%% file: main.tex
\documentclass[sigconf]{acmart}

\copyrightyear{2026}
\acmYear{2026}
\setcopyright{cc}
\setcctype{by}
\acmConference[MM '26]{Proceedings of the 34th ACM International Conference on Multimedia}{November 10--14, 2026}{Rio de Janeiro, Brazil}
\acmBooktitle{Proceedings of the 34th ACM International Conference on Multimedia (MM '26), November 10--14, 2026, Rio de Janeiro, Brazil}
\acmDOI{10.1145/3767308.3836174}
\acmISBN{979-8-4007-2213-4/2026/11}
\usepackage{float}
\usepackage{longtable}
\usepackage{amsmath}
\usepackage{algorithm}
\usepackage{algpseudocode}
\usepackage{booktabs}
\usepackage{makecell}
\usepackage{multirow}
\usepackage{siunitx}
\usepackage{graphicx}
\usepackage{subcaption}
\usepackage[utf8]{inputenc}
\usepackage{latexsym}
\usepackage[export]{adjustbox}
\usepackage[T1]{fontenc}
\usepackage{microtype}
\usepackage{array}
\usepackage{ragged2e}
\makeatletter
\newcommand{\ContIndent}{\hspace{\dimexpr\ALG@thistlm+\algorithmicindent\relax}}
\makeatother

\begin{document}

\title{Text Steganography with Dynamic Codebook and \\Multimodal Large Language Model}

\author{Jianxin Gao}
\orcid{0009-0006-3297-4792}
\email{jxgao@cau.edu.cn}
\affiliation{%
  \department{College of Information and Electrical Engineering}
  \institution{China Agricultural University}
  \city{Beijing}
  \country{China}
}

\author{Ruohan Lei}
\orcid{0009-0003-4159-6402}
\email{lrh07@cau.edu.cn}
\affiliation{%
  \department{College of Information and Electrical Engineering}
  \institution{China Agricultural University}
  \city{Beijing}
  \country{China}
}

\author{Wanli Peng}
\orcid{0000-0001-9636-6928}
\authornote{Corresponding author.}
\email{wlpeng@cau.edu.cn}
\affiliation{%
  \department{College of Information and Electrical Engineering}
  \institution{China Agricultural University}
  \city{Beijing}
  \country{China}
}

\renewcommand{\shortauthors}{Jianxin Gao, Ruohan Lei, \& Wanli Peng}

\begin{abstract}
With the popularity of the large language models (LLMs), text steganography has achieved remarkable performance.
However, existing methods still have some issues:
(1) For the white-box paradigm, this steganography behavior is prone to exposure due to sharing the off-the-shelf language model between Alice and Bob.
(2) For the black-box paradigm, these methods lack flexibility and practicality since Alice and Bob should share the fixed codebook while sharing a specific extraction prompt for each steganographic sentence. 
In order to improve the security and practicality,
we introduce a black-box text steganography with a dynamic codebook and multimodal large language model.
Specifically,
we first construct a dynamic codebook via some shared session configuration and a multimodal large language model.
Then an encrypted steganographic mapping is designed to embed secret messages during the steganographic text generation.
Furthermore, we introduce a feedback optimization mechanism based on reject sampling to ensure accurate extraction of secret messages.
Experimental results show that the proposed method outperforms existing white-box text steganography methods in terms of embedding capacity and text quality.
Meanwhile, the proposed method has achieved better practicality and flexibility than the existing black-box paradigm in some popular online social networks.
\end{abstract}




\maketitle

\input{sections/Introduction}
\input{sections/RelatedWork}
\input{sections/method}
\input{sections/experiments}
\input{sections/video}

\section{Conclusion}
In this paper, we presented DyCo-Stega for black-box text steganography with a dynamic codebook and multimodal large language models. 
The core idea is to use public visual context to construct a dynamic codebook for each communication session so that secret extraction does not rely on fixed codebooks. 
Under this framework, the secret message can be extracted by reconstructing the dynamic codebook and recovering the offset with the transmitted private key and the session seed. 
In addition, the optimization procedures for image generation and caption generation improve the reliability of seed word recovery and secret extraction in black box settings.
Extensive experiments demonstrate the effectiveness of the proposed method. 
DyCo-Stega achieves strong overall performance in text quality, embedding capacity, statistical imperceptibility, and anti-steganalysis ability. 
The results also show that the method preserves image--text consistency, remains robust across different generators and recognizers, and maintains high recovery performance after transmission through online social networks. 
We further provide a preliminary study on video carriers, which suggests that the proposed framework can be extended beyond images. 
In future work, we will investigate the feasibility of applying the proposed framework to other modalities, such as audio, and further study the use of multimodal large language models for more stable steganography.

\begin{acks}
This work was supported by the National Natural Science Foundation of China (No.~62402117).
\end{acks}

\newpage
\bibliographystyle{ACM-Reference-Format}
\bibliography{ref} 
\appendix
\input{sections/appendix}

\end{document}

%% file: sections/Introduction.tex
\section{Introduction}

Text steganography is an active research topic in the fields of information hiding and neural language generation (NLG). 
It is the art of embedding secret information into innocent text data, aiming to covertly transmit secret information between Alice (sender) and Bob (receiver) \cite{anderson1998limits,provos2003hide,cox2007digital}.
In the steganography field, the innocent text is named as \textit{cover text}.
The generated text with the secret message is called \textit{stego text}.
Due to its inherent robustness, text data has become an appropriate carrier for steganography in online social networks (OSNs), such as Instagram, X, Weibo, and WeChat~\cite{ziegler2019neural,zhang2021adg}.
Early text steganography mainly relied on modifying existing text, for example, through synonym substitution or controlled rewriting~\citep{chang2010paraphrases,chang2012wordorder,ziegler2019neural}. Although these methods can preserve part of the original meaning, they usually provide limited payload and are often vulnerable to linguistic or statistical detection~\citep{chang2014practical,ueoka2021masked,chen2011relative,xiang2018wordembedding,wen2019cnn,yang2019tsrnn,wu2021gnn}.

In the past decades, language models based on deep learning have achieved superior performance in a variety of neural language generation tasks.
These kinds of tasks aim to generate realistic and plausible textual content that is indistinguishable from human-written text, which opens up a new research avenue for highly secure text steganography, namely, \emph{generative text steganography (GTS)} 
~\cite{fang2017generating,yang2019rnnstega,zhang2021adg,kaptchuk2021meteor,schroeder2022perfectly,ding2023discop,wang2025sparsamp,huang-etal-2026-od,lin-etal-2024-zero,yan-murawaki-2025-addressing}.
The paradigm of GTS can be formulated as follows:
\begin{equation}
    T_{s}=f_{LM}\left[T_{p}, k, g\left(m\right)\right]
\end{equation}
where $T_{s}$ is generated stego text. 
$f_{LM}[\cdot]$ represents an off-the-shelf language model. 
$T_{p}$ denotes a prefix or context. 
$k$ is a private key shared between Alice and Bob. 
$m$ is the secret message.
$g(\cdot)$ denotes a steganographic mapping function that is elaborately designed to embed secret messages.
The difficulties of GTS  mainly come from $f_{LM}[\cdot]$ and $g(\cdot)$. 
The former inclines to cause quality reduction of stego text since most GTS methods generate stego texts in the inference phase of the off-the-shelf language models.
The latter mainly impacts the security of the GTS method, in particular anti-steganalysis and statistical imperceptibility.
In the GTS task, the main challenge is how to make a satisfying trade-off between text quality and security.
It is obvious that using a better language model $f_{LM}[\cdot]$ can significantly boost the semantic consistency and fluency of generated stego texts since GTS can be regarded as a special text generation task under the control of secret messages.

Although existing generative text steganographic methods substantially improved the trade-off among embedding capacity, text quality, and statistical security,
they have a behavior security risk due to the white-box setting.
Namely, in practice, Alice and Bob require sharing the off-the-shelf language model, which is prone to exposing the steganographic behavior since the deployment of a shared language model, especially the large language model, is a special behavior.
In order to enhance the behavior security, Wu et al.~\cite{wu2024llmstega} proposed an LLM-Stega based on prompt engineering.
The main goal of this method is concealing the steganographic behavior with the common usages of large language models.
However, the black-box method lacks flexibility and practicality since it relies heavily on a shared codebook, and the correct extraction of different stego sentences requires different extraction prompts. 

To address these problems, we propose a black-box text steganography method with a dynamic codebook and a multimodal large language model (MLLM), named DyCo-Stega. 
The key idea is to use public visual context to support the construction of a dynamic codebook so that secret extraction no longer relies on a fixed shared codebook across different communication rounds. 
Under this framework, the MLLM serves not only as a generation interface but also as a synchronization tool between the sender and the receiver~\cite{alayrac2022flamingo,li2023blip2,liu2023visual,bai2025qwen25vl}. In addition, we design a feedback optimization mechanism to improve the reliability of codebook reconstruction and secret extraction in the black-box setting. 
Experimentally, the proposed method offers satisfactory practicality and flexibility while preserving strong text quality and embedding capacity.

The main contributions of this paper are as follows:
\begin{enumerate}
    \item We propose a black-box text steganography framework that constructs a dynamic codebook for each communication round from public image and caption context, rather than relying on a fixed reusable codebook.
    \item We design a complete embedding and extraction process that combines dynamic codebook construction, encrypted message mapping, and a feedback optimization mechanism to support accurate recovery in a black-box setting.
    \item We conduct extensive experiments to evaluate text quality, embedding capacity, statistical imperceptibility, resistance to steganalysis, and robustness in realistic online settings. The results show that the proposed method achieves a strong balance between practicality and performance.
\end{enumerate}

%% file: sections/RelatedWork.tex
\section{Related Work}

\paragraph{White-box linguistic steganography.}
Early text steganography mainly hid messages by editing existing text, such as synonym substitution or controlled rewriting~\citep{chapman1997hiding,chang2014practical}. More recent edit-based work revisited this paradigm with masked language models, showing that neural editing can improve the trade-off between payload and imperceptibility~\citep{ueoka2021masked}. With the development of neural language models, research further shifted toward generative text steganography, where secret bits are embedded during token generation. Representative examples include RNN/LSTM-based methods~\citep{fang2017generating,yang2019rnnstega} and coding-based methods such as arithmetic coding and its variants~\citep{ziegler2019neural,dai2019near}, which substantially improved the trade-off between text quality and embedding capacity. Another major line of work pursues provable security. ADG~\citep{zhang2021adg} embeds messages by adaptively grouping candidate tokens into approximately equal-probability sets, while later methods such as Discop, minimum-entropy-coupling-based steganography, METEOR, and SparSamp~\citep{ding2023discop,schroeder2022perfectly,kaptchuk2021meteor,wang2025sparsamp} further aim to preserve the original sampling distribution more faithfully. Despite their differences, these methods generally assume white-box access to token probabilities, and often require the communicating parties to share the same local generative model, which easily exposes the steganographic behavior.

\paragraph{Black-box linguistic steganography.}
As closed-source LLM services become increasingly prevalent, recent work has begun to explore black-box linguistic steganography under API or user-interface access. In this setting, the design focus shifts from direct control over token probabilities to protocol coordination through prompts and model outputs. Recent examples include zero-shot generative linguistic steganography based on in-context learning~\citep{lin-etal-2024-zero} and LLM-Stega~\citep{wu2024llmstega}, which performs embedding and extraction through pre-shared keyword sets, encrypted mappings, and rejection sampling. More recently, tokenization inconsistency has been identified as an important robustness issue in LLM-based steganography, showing that reliable recovery depends not only on text quality but also on whether sender and receiver can maintain a stable hidden protocol under mismatched tokenization~\citep{yan-murawaki-2025-addressing}. These studies improve practical deployability by removing the need for logit access. However, existing black-box methods still rely heavily on reusable protocol materials, such as fixed lexical inventories, static cross-session mappings, or other persistent local artifacts. By contrast, our method reconstructs the decoding structure on demand for each session from a one-time private convention and public multimodal context, rather than depending on persistent reusable local artifacts.

\paragraph{Multimodal large language models.}
Recent multimodal large language models (MLLMs) have demonstrated strong capabilities in image understanding, visual grounding, and instruction-following caption generation, making them practical black-box interfaces for multimodal interaction~\citep{alayrac2022flamingo,li2023blip2,dai2023instructblip,liu2023visual,bai2025qwen25vl}. These capabilities are directly relevant to our setting: visual grounding supports reliable recovery of image-side semantic anchors, instruction following supports constrained caption generation, and image-conditioned generation allows public visual content to serve as a shared synchronization substrate. However, existing MLLM research has primarily focused on perception and generation tasks, rather than covert communication. To the best of our knowledge, prior black-box linguistic steganography methods have not used public image--caption context to reconstruct a session-specific codebook. Our work fills this gap by turning multimodal public context into a session-level synchronization interface.

%% file: sections/method.tex
\section{The Proposed Methodology}
\label{sec:method}

% ---- lightweight local macros (optional; keep local to Method) ----
\providecommand{\Tok}{\operatorname{Tok}}
\providecommand{\Def}{\operatorname{Def}}
\providecommand{\Syn}{\operatorname{Syn}}
\providecommand{\Ex}{\operatorname{Ex}}
\providecommand{\Freq}{\operatorname{Freq}}
\providecommand{\Root}{\operatorname{Root}}
\providecommand{\Norm}{\operatorname{Norm}}
\providecommand{\Dedup}{\operatorname{Dedup}}

% ---- consistent math subscripts (optional) ----
\providecommand{\org}{\mathrm{org}}
\providecommand{\ext}{\mathrm{ext}}
\providecommand{\dyn}{\mathrm{dyn}}
\providecommand{\sem}{\mathrm{sem}}
\providecommand{\gen}{\mathrm{gen}}
\providecommand{\emb}{\mathrm{emb}}
\providecommand{\edit}{\mathrm{edit}}
\providecommand{\tar}{\mathrm{tar}}
\providecommand{\ord}{\mathrm{ord}}

% string concatenation (token-string concatenation with a fixed delimiter)
\providecommand{\cat}{\mathbin{\Vert}}

\providecommand{\Hash}{\mathcal{H}} % cryptographic hash (e.g., SHA-256)
\providecommand{\HtwoI}{\operatorname{Hash2Int}}
\providecommand{\IntToBin}{\operatorname{Int2Bin}}
\providecommand{\BinToInt}{\operatorname{Bin2Int}}

% Concatenation (bitstrings / encodings)
\providecommand{\concat}{\mathbin{\|}}
\providecommand{\coloneqq}{\mathrel{\mathop:}=}% fallback if mathtools is not loaded

\newcommand{\strcat}{\mathbin{\Vert}} % string concatenation
\newcommand{\ser}{\operatorname{ser}} % canonical serialization to a string
\newcommand{\PSK}{\mathit{PSK}}
\providecommand{\AncSeq}{\textit{AncSeq}}

\begin{figure*}[t]
    \centering
    \includegraphics[width=\textwidth]{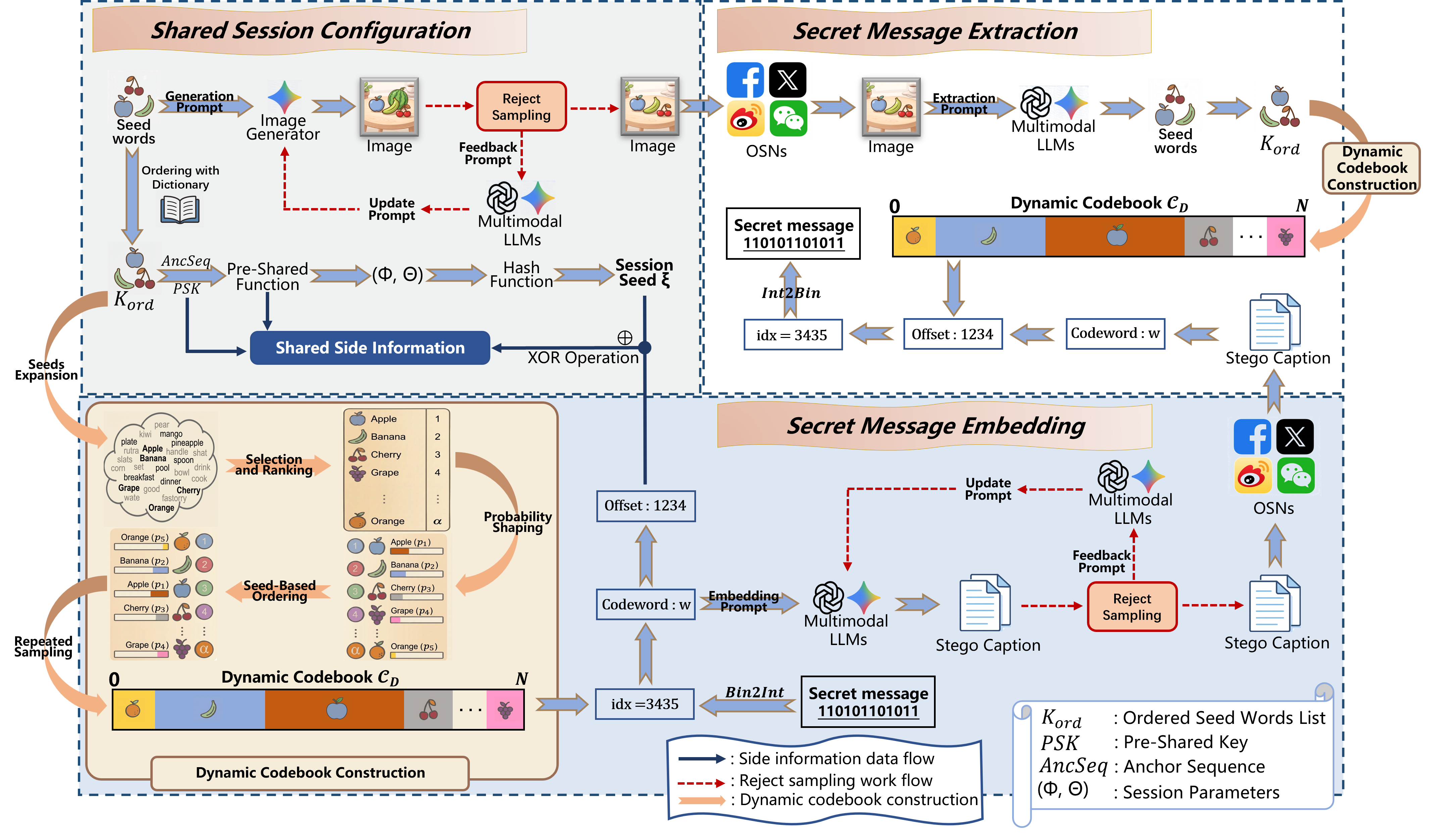}
    \captionsetup{font=bf}
    \caption{The overall framework of DyCo-Stega. The figure shows the three main steps. Crucially, the dynamic codebook construction is executed independently by both parties without transmitting the codebook itself.}
    \label{fig:overview}
\end{figure*}

As shown in Figure~\ref{fig:overview}, the proposed DyCo-Stega framework consists of three main stages: shared session configuration, secret message embedding, and secret message extraction. 
In practice, Alice leverages a multimodal large language model to generate an image with a specific semantic controlled by the seed words. 
Then these seed words are used to construct a dynamic codebook for mapping the secret message bits while generating a steganographic sentence.
The generated semantic image and steganographic text are sent to Bob via some OSNs.
Finally, Bob uses the semantic image and side information to reconstruct the same codebook and then correctly extracts the secret messages.

\subsection{Shared Session Configuration}
\label{subsec:sync}
This stage aims to establish a shared session configuration, which is shared with Bob for the correct extraction of secret messages.
The configuration contains a pre-shared key $\PSK$, the ordered seed words $\mathcal{K}_{ord}$, and an anchor sequence $\AncSeq$.
Specifically, an image is generated by using the seed words and an MLLM with a reject sampling mechanism, where the generation prompt should be updated iteratively to ensure Bob can extract the same seed words. 
The scene description and the seed words are used to construct a generation prompt.
The generated image, together with an extraction prompt, is fed into a multimodal LLM to extract a set of seed words, denoted by $\mathcal{K}_{\ext}$. 
If $\mathcal{K}_{\ext}$ does not exactly match the seed words, an image feedback prompt is used to analyze the discrepancy and produce an update prompt for image refinement. This reject sampling procedure continues until the extracted seed words exactly match the original seed words.
The algorithm detail is shown in Algorithm~\ref{alg:ev_rs}.

\begin{algorithm}[t]
\caption{Image generation based on reject sampling}
\label{alg:ev_rs}
\begin{algorithmic}[1]
\Require Scene description, seed words, generation prompt template, extraction prompt, image feedback prompt template, image generator, multimodal LLM
\Ensure Image $I$

\State $\mathit{accepted} \gets \mathrm{False}$
\State $P^{\gen} \gets$ construct generation prompt using generation prompt template, scene description, and seed words
\State $I \gets$ generate an image using image generator with $P^{\gen}$

\While{not $\mathit{accepted}$}
    \State $\mathcal{K}_{\ext} \gets$ extract seed words from image $I$ using multimodal LLM with extraction prompt
    \If{$\mathcal{K}_{\ext} = \text{seed words}$}
        \State $\mathit{accepted} \gets \mathrm{True}$
    \Else
        \State $P^{\mathrm{fb}}_{I} \gets$ construct image feedback prompt using image feedback prompt template, seed words, and $\mathcal{K}_{\ext}$
        \State $P^{\mathrm{upd}}_{I} \gets$ get update prompt using multimodal LLM with image $I$ and $P^{\mathrm{fb}}_{I}$
        \State $I \gets$ refine image $I$ using image generator with $P^{\mathrm{upd}}_{I}$
    \EndIf
\EndWhile

\If{$\mathit{accepted}$}
    \State \Return Image $I$
\EndIf
\end{algorithmic}
\end{algorithm}

Then, the seed words are sorted deterministically according to their word frequencies in a public dictionary $\mathcal{D}$, yielding the ordered seed word list $\mathcal{K}_{\ord}=(k_1,\dots,k_K)$. 
Using $\mathcal{K}_{\ord}$, a pre-shared key $\PSK$, and a public anchor sequence $\AncSeq$ (e.g., punctuation patterns or emojis), the session parameters are computed as follows:
\begin{equation}
\label{eq:mapping}
(\Phi, \Theta) = g(\PSK, \mathcal{K}_{\ord}, \AncSeq),
\end{equation}
where $\Phi=(\alpha,b,\sigma)$ contains the codebook size $\alpha$, the bit capacity per codeword $b$, and the shaping parameter $\sigma$. 
The parameter $\Theta$ is introduced to increase the unpredictability of the session seeds.  
$\AncSeq$ is specified by the sender and can be directly extracted from the OSNs. 
Both the $\PSK$ and the function $g$ are established via a one-time private agreement. 

Finally, an integer session seed $\xi$ is derived from $\Phi$ and $\Theta$ using a cryptographic hash function $\Hash(\cdot)$ (e.g., SHA-256) and an integer mapping function $\HtwoI(\cdot)$:
\begin{equation}
\label{eq:seed}
\xi = \HtwoI\bigl(\Hash(\alpha \cat b \cat \sigma \cat \Theta)\bigr).
\end{equation}
The seed $\xi$ determines the ordering during codebook construction and masks the within-interval offsets during secret message embedding.
\subsection{Secret Message Embedding}
\label{subsec:embed}

Given the above shared session configuration, we first construct the dynamic codebook $\mathcal{C}_{D}$ and then map the secret message to a codeword $\tilde{w}_j$ in $\mathcal{C}_{D}$ together with an interval offset.

\textbf{Dynamic Codebook Construction}
The dynamic codebook spans the integer range $[0,N)$, where $N=2^b$. 
The construction consists of five stages: seed expansion, selection and ranking, probability shaping, seed-based ordering, and repeated sampling. We summarize the overall process  in Algorithm \ref{alg:dyn_codebook}.

\begin{algorithm}[t]
\caption{Dynamic codebook construction}
\label{alg:dyn_codebook}
\begin{algorithmic}[1]
\Require Ordered seed word list $\mathcal{K}_{ord}=\{k_1,\dots,k_K\}$, public dictionary $\mathcal{D}$, codebook size $\alpha$, bit capacity per codeword $b$, shaping parameter $\sigma$, session seed $\xi$, reverse-definition weight $\lambda_{\mathrm{rev}}$, forward-context weight $\lambda_{\mathrm{fwd}}$, prefix weight $\lambda_{\mathrm{pre}}$
\Ensure Dynamic codebook $\mathcal{C}_D$

\State $N \gets 2^b$, \quad $\mathrm{score}(w) \gets 0,\ \forall w \in \mathcal{D}$

\ForAll{$k \in \mathcal{K}$}
    \ForAll{$w \in \mathcal{D}$}
        \State if $k$ appears in the definition text of $w$:
        \Statex \ContIndent $\mathrm{score}(w) \gets \mathrm{score}(w)+\lambda_{\mathrm{rev}}$
        \State if $w$ appears in the definition text, example sentences, or synonyms of $k$:
        \Statex \ContIndent $\mathrm{score}(w) \gets \mathrm{score}(w)+\lambda_{\mathrm{fwd}}$
        \State if $w$ has prefix $k$:
        \Statex \ContIndent $\mathrm{score}(w) \gets \mathrm{score}(w)+\lambda_{\mathrm{pre}}$
    \EndFor
\EndFor

\State $\mathcal{V}_{\mathrm{cand}} \gets$ the top $\alpha-K$ non-seed words ranked by $\mathrm{score}(w)$
\State sort $\mathcal{V}_{\mathrm{cand}}$ by descending dictionary frequency
\State $\mathcal{V}_{\mathrm{sem}} \gets$ place $\mathcal{K}_{ord}$ in the first $K$ positions, followed by $\mathcal{V}_{\mathrm{cand}}$

\State computes weight: $q_j \gets \exp\!\left(-\frac{(j-1)^2}{2\sigma^2}\right) ,\ \forall j=1,\dots,\alpha $

\State obtain the final probability: $p_j \gets \dfrac{q_j}{\sum_{i=1}^{\alpha} q_i}$

\State permute $\{(w_j,p_j)\}_{j=1}^{\alpha}$ into $\{(\tilde{w}_j,\tilde{p}_j)\}_{j=1}^{\alpha}$

\State repeated sampling: $L_j \gets \tilde{p}_j \cdot N,\ \forall j=1,\dots,\alpha$
\State $B_0 \gets 0,\quad B_j \gets \sum_{i=1}^{j} L_i,\quad \Omega_j \gets [B_{j-1}, B_j),\ \forall j=1,\dots,\alpha$
\State \Return $\mathcal{C}_D = \{(\tilde{w}_j,\Omega_j)\}_{j=1}^{\alpha}$

\end{algorithmic}
\end{algorithm}

\paragraph{(1) Seeds Expansion.}
DyCo-Stega expands the seed words into a broader candidate set using a public dictionary $\mathcal{D}$. 
Specifically, candidate retrieval follows three dictionary-based routes. 
\emph{Reverse-definition matching} retrieves candidate words whose dictionary definitions contain a seed word. 
\emph{Forward-context matching} retrieves candidate words that appear in the dictionary entry of a seed word, including its definition text, example sentences, and synonyms. 
\emph{Prefix matching} retrieves candidate words that use a seed word as their prefix.
Each route contributes a fixed weight, and the final score of a candidate word is obtained by summing the weights of all matched routes. This dictionary-based expansion requires neither local LLM weights nor embedding models.

\paragraph{(2) Selection and Ranking.}
To construct the final semantic pool $V_{\text{sem}}$, we retain all seed words and select the $\alpha - K$ highest-scoring candidates. 
The selected words are then divided into two groups: the seed words and the remaining $\alpha - K$ words. Within each group, words are independently sorted in descending order according to their dictionary frequencies. 
This procedure ensures that $V_{\text{sem}}$ is deterministic while prioritizing the seed words and high-frequency words.

\paragraph{(3) Probability Shaping.}
To improve the linguistic quality of the generated stego captions, we assign a probability distribution that gives higher-ranked words larger probabilities in $V_{\sem}$. 
Since the words in $V_{\sem}$ are already sorted by descending dictionary frequency, this design makes seed words and high-frequency words more likely to be selected. 
Let $j \in \{1, \dots, \alpha\}$ denote the rank of the word $w_j$ in $V_{\sem}$, and let $\sigma > 0$ be the shaping parameter derived from the shared session configuration. 
A half-Gaussian kernel function computes the unnormalized weight $q_j$ for each rank, which is then normalized to obtain the final probability $p_j$:
\begin{equation}
\label{eq:rj}
q_j = \exp\!\left(-\frac{(j-1)^2}{2\sigma^2}\right), \qquad
p_j = \frac{q_j}{\sum_{i=1}^{\alpha} q_i}.
\end{equation}

\paragraph{(4) Seed-Based Ordering.}
Using the session seed $\xi$ to initialize a pseudo-random number generator (PRNG), we reorder the original word-probability pairs $(w_j, p_j)$ into a new sequence $(\tilde{w}_j, \tilde{p}_j)$. This reordering changes only the position of each word within the sequence, and every word strictly retains its assigned probability. Consequently, the final codebook layout is dependent on the session seed $\xi$, thereby enhancing the security of the dynamic codebook.

\paragraph{(5) Repeated Sampling.}
The reordered word-probability pairs are finally converted into contiguous intervals over $[0,N)$. Each word $\tilde{w}_j$ is assigned an interval length $L_j$ according to its reordered probability $\tilde{p}_j$:
\begin{equation}
\label{eq:Lj}
L_j = \tilde{p}_j \cdot N, \qquad \sum_{j=1}^{\alpha} L_j = N.
\end{equation}
Starting from $B_0=0$, these intervals are placed consecutively, and the boundary of the $j$-th interval is computed as
\begin{equation}
\label{eq:boundaries}
B_j = \sum_{i=1}^{j} L_i, \qquad \Omega_j = [B_{j-1},\, B_j).
\end{equation}
The final dynamic codebook is therefore defined as
\begin{equation}
\label{eq:cdyn}
\mathcal{C}_{D} = \bigl\{(\tilde{w}_j,\Omega_j)\bigr\}_{j=1}^{\alpha}.
\end{equation}

\textbf{Encrypted Steganographic Mapping}
After constructing the dynamic codebook $\mathcal{C}_{D}$, we convert the secret message $m$ into an integer index
\[
\mathrm{idx} = \BinToInt(m), \qquad \mathrm{idx}\in[0,N).
\]
We then locate the target interval $\Omega_j=[B_{j-1},B_j)$ that contains this index and select the corresponding word $\tilde{w}_j$ as the codeword. At the same time, we compute the offset
\begin{equation}
\label{eq:offset}
o = \mathrm{idx} - B_{j-1}, \qquad 0 \le o < L_j.
\end{equation}
The codeword identifies the selected codebook interval, while the offset specifies the exact position of the message within that interval.

Directly transmitting the offset would expose the precise location of the message inside the selected interval. To avoid this risk, the offset is not sent through the Online Social Network (OSN). Instead, we protect it with an XOR-based masking strategy. Specifically, we XOR the binary form of the offset with the session seed $\xi$ to obtain a private key $S$:
\begin{equation}
\label{eq:aux_key}
S = \IntToBin(o) \oplus \IntToBin(\xi),
\end{equation}
where $\oplus$ denotes the bitwise XOR operation. This private key $S$ is shared through a side channel. Without the session seed $\xi$, a warden cannot correctly recover the offset even if $S$ is intercepted.

\textbf{Stego Caption Generation}
After determining the codeword, we use an MLLM to generate a stego caption $C$ for the image $I$. For reliable extraction, the generated caption must satisfy three conditions: it must contain the selected codeword $w$, it must contain the anchor sequence $\AncSeq$, and it must not contain any other words from the semantic pool $V_{\sem}$.
The caption generation with a multimodal LLM does not always satisfy all of these constraints. 
Similar to image generation, we also introduce a reject sampling mechanism for caption generation.
First, an embedding prompt guides the multimodal LLM to generate a stego caption $C$. 
If $C$ fails verification, a caption feedback prompt asks the LLM to identify the errors and produce an update prompt for revision. This loop continues until the caption passes verification. 
The accepted stego caption is then posted on the OSN.

\subsection{Secret Message Extraction}
\label{subsec:extraction}
Upon receiving the stego caption $C$, the receiver extracts the codeword $w$ and the anchor sequence $\AncSeq$ from the caption and uses an MLLM to recover the seed words from the shared image $I$. 
The recovered seed words are sorted according to their dictionary frequencies to obtain the ordered seed word list $\mathcal{K}_{\ord}$. 
Using $\mathcal{K}_{\ord}$, $\AncSeq$, and the pre-shared key $\PSK$, the receiver recomputes the session parameters and the session seed $\xi$, and then reconstructs the dynamic codebook $\mathcal{C}_{D}$.

To recover the exact integer index of the secret message, the receiver first restores the offset $o$ by performing an XOR operation between the private key $S$ (received via the side channel) and the binary form of the session seed $\xi$:
\begin{equation}
\label{eq:decode_offset}
\IntToBin(o) = S \oplus \IntToBin(\xi).
\end{equation}
Next, the receiver locates the target interval $\Omega_j=[B_{j-1},B_j)$ corresponding to the codeword $w$ in the codebook $\mathcal{C}_{D}$. 
The index is computed by adding the offset $o$ to the base boundary $B_{j-1}$:
\begin{equation}
\label{eq:decode_idx}
\mathrm{idx} = B_{j-1} + o.
\end{equation}
Finally, the integer index $\mathrm{idx}$ is converted back to its binary representation to recover the original secret message $m$.

%% file: sections/experiments.tex
\section{Experiments}
\label{sec:experiments}

% ---- local metric macros (notation) ----
\providecommand{\PPL}{\ensuremath{\mathrm{PPL}}}
\providecommand{\SS}{\ensuremath{\mathrm{SS}}}
\providecommand{\KLD}{\ensuremath{\mathrm{KLD}}}
\providecommand{\EC}{\ensuremath{\mathrm{EC}}}
\providecommand{\Acc}{\ensuremath{\mathrm{Acc}}}

% (Dependencies: booktabs, makecell, multirow, siunitx)
% ==================================================================
\subsection{Experimental Setup}

\noindent\textbf{(1) Baselines.}
We compare DyCo-Stega with one black-box baseline, LLM-Stega~\citep{wu2024llmstega}, and six advanced white-box baselines: AC~\citep{ziegler2019neural}, ADG~\citep{zhang2021adg}, Discop~\citep{ding2023discop}, SparSamp~\citep{wang2025sparsamp}, METEOR~\citep{kaptchuk2021meteor}, and iMEC~\citep{schroeder2022perfectly}.

\noindent\textbf{(2) Models and Parameters.}
 We use the gemini-3-pro-image-preview API for image generation and the gpt-5.1-chat-latest API for stego caption generation. 
 We set $(\alpha,b,\sigma)=(24,28,2.5)$ and embed $\tau{=}5$ codewords in each stego caption. 
 For the LLM-Stega, we also use the gpt-5.1-chat-latest API to generate stego texts. 
 Since white-box methods require access to token probabilities, we implement them on GPT-2~\citep{radford2019gpt2} and Qwen2.5~\citep{yang2024qwen25}. 
 For DyCo-Stega and LLM-Stega, we generate 15,000 stego texts. For each white-box baseline, we select appropriate temperatures and set top-$p{=}0.99$, and generate 10,000 stego texts on each of the two language models, resulting in a total of $6\times2\times10{,}000$ stego texts.
% ==================================================================

\subsection{Evaluation Metrics}
\noindent\textbf{(1) Text Quality.}
We use Perplexity (PPL) and Semantic Similarity (SS) to evaluate the text quality of generated stego texts. PPL is a widely used quantitative metric in text generation tasks. We compute the PPL of different stego texts using GPT-2~\citep{radford2019gpt2} as the reference language model:
\begin{equation}
\mathrm{PPL}=\exp\left(
-\frac{1}{L}\sum_{t=1}^{L}\log p(w_t|w_{<t})
\right),
\end{equation}
where $L$ is the length of the text, $w_t$ is the $t$-th token, and $p(w_t|w_{<t})$ is the conditional probability assigned by the language model to token $w_t$ given the preceding context. Lower PPL indicates better fluency.

For semantic similarity, we adopt Sentence-BERT~\citep{reimers2019sbert} and use the roberta-base-nli-mean-tokens~\citep{liu2019roberta} model to extract sentence embeddings. We then compute the average cosine similarity between cover and stego texts. Higher SS indicates better semantic preservation after secret message embedding.

\noindent\textbf{(2) Embedding Capacity.}
Embedding Capacity (EC) is the average number of secret bits embedded per word, which is measured in bits per word (bpw). It is calculated as
\begin{equation}
\mathrm{EC}=\frac{B}{W}\quad (\mathrm{bpw}),
\end{equation}
where $B$ is the number of embedded bits and $W$ is the total number of words in the stego texts.

\noindent\textbf{(3) Anti-Steganalysis Ability.} 
Anti-steganalysis ability is important for security. We use the test accuracy of a steganalysis classifier based on BERT-FT~\citep{peng2021realtime}. For each method, we construct a balanced cover--stego dataset and divide it into 60\%/20\%/20\% for training, validation, and testing. The steganalysis accuracy is defined as 
\begin{equation}
\mathrm{Acc}=\frac{TP+TN}{TP+TN+FP+FN},
\end{equation}
where $TP$ is true positives, $TN$ is true negatives, $FP$ is false positives, and $FN$ is false negatives. An accuracy closer to $0.5$ indicates weaker detectability and thus better concealment. 

\noindent\textbf{(4) Statistical Imperceptibility.}
The Kullback-Leibler Divergence (KLD) is used to measure the statistical imperceptibility by comparing the distribution of stego text embeddings with that of cover text embeddings. We select the KLD metric proposed by Zhang et al.~\citep{zhang2021adg} to evaluate statistical imperceptibility:
\begin{equation}
KLD=\sum\left[\log\left(\frac{\sigma_y}{\sigma_x}\right)+\frac{\sigma_x^2+(\mu_x-\mu_y)^2}{2\sigma_y^2}-\frac{1}{2}\right]
\end{equation}
where $\mu_x$ and $\sigma_x$ are the mean and standard deviation of cover texts, while $\mu_y$ and $\sigma_y$ represent those of stego texts.

\noindent\textbf{(5) Image-Text Consistency.} 
We use CLIP-Score~\citep{hessel2021clipscore}, computed with CLIP ViT-B/32~\citep{radford2021clip}, to evaluate the semantic alignment between the image and the generated caption. Higher CLIP-Score indicates better image-text consistency.

\noindent\textbf{(6) Human Evaluation.}
For human evaluation, we invited 30 computer science undergraduates to assess the generated stego texts. We randomly sampled and anonymized stego texts generated by DyCo-Stega, LLM-Stega, and six white-box baselines before annotation. Each annotator independently rated the captions on a 1--5 scale across three dimensions: fluency, clarity, and grammaticality.
% ==================================================================
\subsection{Main Results}
\label{sec:main_results}

\noindent\textbf{(1) Main Results.}
Table~\ref{tab:main_results} summarizes the main benchmark results in text quality, embedding capacity, statistical imperceptibility, and anti-steganalysis ability. Among all evaluated methods, DyCo-Stega achieves the best overall performance, with a PPL of 128.9, an SS of 0.8940, a KLD of 3.37, and an EC of 6.37 bpw. Compared with LLM-Stega, our method obtains a lower PPL, higher SS, lower KLD, higher EC, and an Acc value closer to 0.5, indicating a better balance between text quality and embedding capacity. The improvement in SS is mainly due to the introduction of the image as an additional semantic constraint. Compared with the white-box baselines, DyCo-Stega also outperforms the best reported values on PPL, SS, KLD, and EC. Note that the KLD metric is computed in the sentence-embedding space rather than the token-probability space; therefore, distribution-preserving methods such as Discop, whose theoretical guarantee holds in the token-probability space, still yield non-trivial KLD values in Table~\ref{tab:main_results}. In particular, although SparSamp on GPT-2 shows slightly stronger anti-steganalysis ability, its text quality is still significantly worse than that of the two black-box methods, as reflected by its much higher PPL and lower semantic similarity.

\begin{table}[t]
\centering
\caption{Main benchmark comparison across white-box and black-box baselines. For Acc, values closer to 0.5 indicate weaker detectability.}
\label{tab:main_results}
\small
\renewcommand{\arraystretch}{1.1}
\begin{tabular*}{\columnwidth}{@{\extracolsep{\fill}}l r r r r r@{}}
\toprule
Method & PPL$\downarrow$ & SS$\uparrow$ & KLD$\downarrow$ & EC$\uparrow$ & Acc \\
\midrule
\multicolumn{6}{@{}l}{\textit{White-box (GPT-2)}} \\
AC       & 410.0 & 0.6980 & 164.20 & 3.89 & 0.5805 \\
ADG      & 876.9 & 0.7070 &  45.11 & 5.33 & 0.4893 \\
Discop   & 396.4 & 0.5500 &  11.42 & 3.80 & 0.4908 \\
SparSamp & 377.9 & 0.5540 &  10.87 & 3.91 & \textbf{0.5000} \\
METEOR   & 298.3 & 0.5490 &  30.83 & 2.42 & 0.4910 \\
iMEC     & 589.1 & 0.5740 &  76.71 & 3.07 & 0.4928 \\
\midrule
\multicolumn{6}{@{}l}{\textit{White-box (Qwen2.5)}} \\
AC       & 380.0 & 0.4360 &  51.43 & 4.16 & 0.7698 \\
ADG      & 461.3 & 0.6480 &  29.87 & 4.86 & 0.5403 \\
Discop   & 145.3 & 0.6910 &  10.45 & 3.63 & 0.5595 \\
SparSamp & 140.0 & 0.6950 &  10.41 & 3.77 & 0.5150 \\
METEOR   & 243.1 & 0.6800 &  24.57 & 3.72 & 0.5188 \\
iMEC     & 405.9 & 0.6560 &  36.22 & 3.46 & 0.5340 \\
\midrule
\multicolumn{6}{@{}l}{\textit{Black-box (GPT-5.1)}} \\
LLM-Stega & 137.8 & 0.7230 &  5.81 & 5.27 & 0.5073 \\
\textbf{DyCo-Stega} & \textbf{128.9} & \textbf{0.8940} & \textbf{3.37} & \textbf{6.37} & 0.4997 \\
\bottomrule
\end{tabular*}
\end{table}

\noindent\textbf{(2) Image--Text Consistency.}
To evaluate whether secret message embedding affects image–text consistency, we analyzed three indoor scenes: \textit{office desk}, \textit{kitchen counter}, and \textit{bathroom vanity}. For each scene, we generated images along with their corresponding cover and stego captions. We then computed the average CLIP-Scores between the images and their cover captions, as well as between the images and their stego captions. As shown in Table~\ref{tab:clip_results}, both types of captions achieve high CLIP-Scores, indicating strong consistency with the image content. Furthermore, the difference in scores between the cover and stego captions is minimal across all scenes, with an average change of only 0.06. These results demonstrate that secret message embedding has a negligible impact on image–text consistency.

\begin{table}[t]
\centering
\caption{CLIP-Score on three representative scenes.}
\label{tab:clip_results}
\small
\setlength{\tabcolsep}{2pt}
\renewcommand{\arraystretch}{1.10}
\begin{tabular*}{\linewidth}{@{\extracolsep{\fill}} l S[table-format=2.2] S[table-format=2.2] S[table-format=+1.2] @{}}
\toprule
Scene & {\makecell{CLIP\\(Cover)}} & {\makecell{CLIP\\(Stego)}} & {$\Delta$} \\
\midrule
Office Desk     & 32.99 & 33.02 & +0.03 \\
Kitchen Counter & 30.00 & 30.16 & +0.16 \\
Bathroom Vanity & 32.23 & 32.22 & -0.01 \\
\midrule
\textbf{Average} & \textbf{31.74} & \textbf{31.80} & \textbf{+0.06} \\
\bottomrule
\end{tabular*}
\end{table}

\begin{figure}
    \centering
    \includegraphics[width=\linewidth]{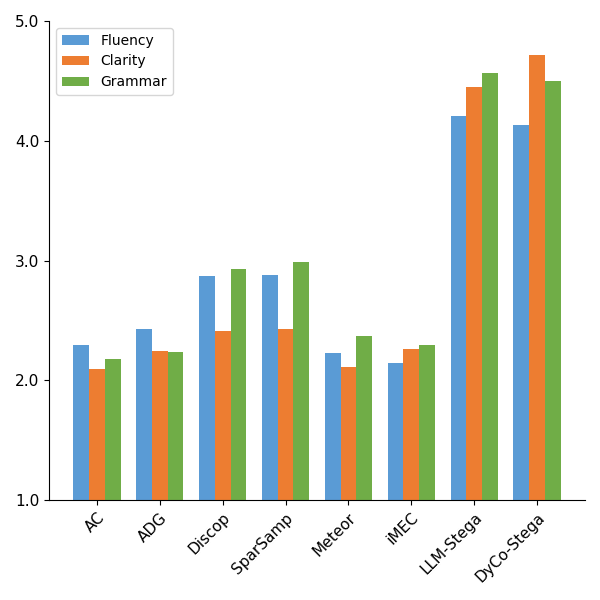}
    \caption{Human evaluation on fluency, clarity, and grammaticality}
    \label{fig:human_eval}
\end{figure}

\noindent\textbf{(3) Human Evaluation.}
The human evaluation results are shown in Figure~\ref{fig:human_eval}. Overall, the two black-box methods obtain substantially higher scores than the white-box baselines on all three dimensions, indicating that black-box text generation produces more natural and readable stego texts. In particular, DyCo-Stega attains the highest score on clarity, while maintaining fluency and grammaticality at a level comparable to LLM-Stega. By contrast, white-box methods achieve much lower overall scores. This is primarily because the open-weight models they rely on are generally less capable than state-of-the-art black-box models. Additionally, the secret message embedding in these white-box methods more easily degrades the naturalness of the generated text. In summary, DyCo-Stega can preserve good text quality while embedding secret messages.

% ==================================================================
\subsection{Ablation Studies}
\label{sec:ablation}

\begin{table*}[t]
\centering
\caption{Comprehensive ablation studies.
  \textbf{(a)}~Cross-model robustness across generator--recognizer pairs, where Gen.\,=\,image generator and Rec.\,=\,recognizer.
  \textbf{(b)}~Probability shaping ablation ($\star$\,=\,default).
  \textbf{(c)}~Robustness to OSN compression.
  \textbf{(d)}~Effect of prompt design and temperature on recovery rate (S@3);
  Full\,=\,our designed prompt, Mini\,=\,minimally constrained prompt.}
\label{tab:combined-ablation}
\small
\setlength{\tabcolsep}{4pt}
\renewcommand{\arraystretch}{1.05}

\noindent\rule{\textwidth}{0.08em}\par\vspace{2pt}

\noindent
\begin{minipage}[c]{0.36\textwidth}
  \centering\textbf{(a) Cross-Model Robustness}\par\vspace{4pt}
  {\renewcommand{\arraystretch}{1.30}%
  \begin{tabular*}{\linewidth}{@{\extracolsep{\fill}}ll ccc@{}}
  \toprule
  \textbf{Gen.} & \textbf{Rec.} & \textbf{S@1} & \textbf{S@3} & \textbf{AvgR} \\
  \midrule
  \multirow{4}{*}{Gemini}
    & Gemini & 83\% & 98\% & 1.21 \\
    & GPT    & 80\% & 97\% & 1.23 \\
    & Grok   & 74\% & 96\% & 1.28 \\
    & Claude & 81\% & 97\% & 1.25 \\
  \midrule
  \multirow{4}{*}{GPT}
    & Gemini & 76\% & 94\% & 1.31 \\
    & GPT    & 72\% & 93\% & 1.36 \\
    & Grok   & 74\% & 92\% & 1.33 \\
    & Claude & 78\% & 93\% & 1.31 \\
  \midrule
  \multirow{4}{*}{Qwen}
    & Gemini & 60\% & 68\% & 1.24 \\
    & GPT    & 55\% & 65\% & 1.27 \\
    & Grok   & 52\% & 64\% & 1.28 \\
    & Claude & 59\% & 66\% & 1.25 \\
  \end{tabular*}}
\end{minipage}%
\hspace{6pt}%
\vrule width 0.4pt%
\hspace{10pt}%
\begin{minipage}[c]{0.56\textwidth}
  \centering\textbf{(b) Probability Shaping Ablation}\par\vspace{4pt}
  \begin{tabular*}{\linewidth}{@{\extracolsep{\fill}}lccc@{}}
  \toprule
  \textbf{Setting}
    & \textbf{PPL\,$\downarrow$}
    & \textbf{KLD\,$\downarrow$}
    & \textbf{SS\,$\uparrow$} \\
  \midrule
  Half-Gaussian ($\sigma{=}\tau/4$)            & 105.7 & 30.32         & 0.903 \\
  Half-Gaussian ($\sigma{=}\tau/2$)\,$\star$   & 128.9 & \textbf{3.37} & 0.894 \\
  Half-Gaussian ($\sigma{=}\tau$)              & 167.9 & 13.40         & 0.836 \\
  Half-Gaussian ($\sigma{=}2\tau$)             & 179.1 & 19.73         & 0.814 \\
  Half-Gaussian ($\sigma{=}4\tau$)             & 182.4 & 38.21         & 0.854 \\
  \midrule
  Uniform                                      & 247.0 & 12.35         & 0.805 \\
  \bottomrule
  \end{tabular*}

  \vspace{8pt}

  \begin{minipage}[b]{0.46\linewidth}
    \centering\textbf{(c) OSN Robustness}\par\vspace{4pt}
    {\renewcommand{\arraystretch}{1.18}%
    \begin{tabular*}{\linewidth}{@{\extracolsep{\fill}}cccc@{}}
    \toprule
    \textbf{Platform} & \textbf{S@1} & \textbf{S@2} & \textbf{S@3} \\
    \midrule
    WeChat    & 97\% & 100\% & 100\% \\
    Instagram & 98\% & 99\%  & 100\% \\
    Facebook  & 98\% & 100\% & 100\% \\
    X         & 97\% & 99\%  & 100\% \\
    WeiBo     & 98\% & 99\%  & 100\% \\
    \end{tabular*}}
  \end{minipage}%
  \hfill
  \begin{minipage}[b]{0.50\linewidth}
    \centering\textbf{(d) Prompt \& Temperature}\par\vspace{4pt}
    {\renewcommand{\arraystretch}{1.18}%
    \begin{tabular*}{\linewidth}{@{\extracolsep{\fill}}ccc@{}}
    \toprule
    \textbf{Temperature} & \textbf{Full} & \textbf{Mini} \\
    \midrule
    0   & 100\%  & 40.7\% \\
    0.7 & 100\%  & 48.7\% \\
    1.0 & 99.3\% & 39.3\% \\
    1.5 & 94.0\% & 34.7\% \\
    2.0 & 90.8\% & 30.2\% \\
    \end{tabular*}}
  \end{minipage}
\end{minipage}

\par\vspace{2pt}
\noindent\rule{\textwidth}{0.08em}

\vspace{4pt}
% --------------------------------------------------------
\end{table*}

\noindent\textbf{(1) Cross-Model Robustness.}
This experiment evaluates whether the proposed method remains reliable across heterogeneous generator--recognizer combinations, rather than depending on a specific model pair. We use S@k to measure the fraction of trials that correctly recover the seed words within $k$ attempts, and AvgR to report the average number of attempts over successful trials only. As shown in Table~\ref{tab:combined-ablation}(a), \emph{Gemini} as the generator achieves consistently strong performance, reaching 96\%--98\% S@3 across all recognizers, while \emph{GPT} also remains stable at 92\%--94\% S@3. In contrast, \emph{Qwen} yields a substantially lower S@3 of 64\%--68\%, although its AvgR on successful cases remains comparable to the other generators. These results indicate that the proposed method generalizes well across recognizer families, but its reliability still depends on how clearly the generator preserves the target visual semantic anchors.

\noindent\textbf{(2) Probability Shaping Mechanism.}
This ablation examines how the probability shaping parameter $\sigma$ affects the trade-off between text quality and concealment in the dynamic codebook construction. As shown in Table~\ref{tab:combined-ablation}(b), a sharper distribution with $\sigma{=}\tau/4$ gives the lowest PPL (105.7) and the highest SS (0.903), but also leads to a much larger KLD (30.32), indicating a stronger deviation from the cover text distribution. As $\sigma$ increases to $\tau$ and $2\tau$, SS decreases from 0.894 to 0.836 and 0.814, while KLD increases from 3.37 to 13.40 and 19.73. The default setting $\sigma{=}\tau/2$ provides the best overall balance, achieving relatively low PPL and KLD while preserving high semantic similarity. Overall, the uniform setting is clearly less favorable than the default half-Gaussian shaping, which confirms that assigning larger probability to higher-ranked words is important for text quality and statistical imperceptibility.

\noindent\textbf{(3) OSN Robustness.}
This experiment evaluates whether transmission through online social networks affects seed word extraction. We upload 500 images that have already been verified as successful to major social platforms, take screenshots of the published posts, and feed the screenshots to the recognizer for seed word extraction.

As reported in Table~\ref{tab:combined-ablation}(c), all five platforms achieve 100\% success under S@3. Under the stricter S@1 and S@2 criteria, S@1 ranges from 97\% to 98\%, and S@2 ranges from 99\% to 100\%. These results show that the proposed method remains highly robust after transmission through online social networks. They also suggest that the method depends mainly on visual semantic content, rather than on fragile low-level image details that may be affected by platform compression or screenshot capture. Since all test images were verified as successful before upload, the small gaps between S@1 and S@3 are more likely to come from the uncertainty of black-box models than from the transmission process itself.

\noindent\textbf{(4) Prompt Structure and Temperature Sensitivity.}
This study evaluates whether the high recovery rate mainly comes from the structured prompt design, and whether this advantage remains stable under different temperatures. As shown in Table~\ref{tab:combined-ablation}(d), the full prompt is highly reliable across a broad temperature range: S@3 stays at 100\% for $T{=}0$ and $T{=}0.7$, remains 99.3\% at $T{=}1.0$, and is still 90.8\% at $T{=}2.0$. In contrast, the minimally constrained prompt achieves only 40.7\% at $T{=}0$, peaks at 48.7\% at $T{=}0.7$, and drops to 30.2\% at $T{=}2.0$. This large performance gap shows that recovery reliability does not come from model capability alone, but critically depends on the structured prompt. Meanwhile, temperature also has a clear effect on recovery performance, and relatively low temperatures generally lead to more accurate recovery.

\noindent\textbf{(5) Deployment Cost.}
We measure the practical deployment overhead of DyCo-Stega over 500 trials with the main setup (Gemini-3 for image generation, GPT-5.1 for caption generation and rejection sampling). The average end-to-end time is 63.5 seconds per image--caption pair, and the API cost per communication round (sender-side generation plus receiver-side extraction) is approximately \$0.32 based on official pricing.

%% file: sections/video.tex
\section{Extension to Video}
\label{sec:video_extension}

Since our method relies on visual semantic anchors, it is natural to explore whether we can use the video generation and understanding abilities of multimodal large language models to transmit videos instead of images to build the dynamic codebook. However, current video generation is less stable than image generation, and it costs more time and money. Therefore, we only conduct a lightweight exploration using the user interfaces of main models.

\noindent\textbf{Experimental Setup.}
For video generation, we select three models: Veo from Gemini, Sora from GPT, and Seedance from Doubao. We take 100 images generated by Gemini from our previous experiments and use them as base images to generate videos. For video recognition, we select five models that can directly read videos: Gemini 3.1 Flash, Gemini 3.1 Pro, GPT 5.3, GPT 5.4 Thinking in Standard mode, and GPT 5.4 Thinking in Extended mode. We upload the generated videos to these recognizers along with an extraction prompt, and record the success rate within three attempts (S@3).

\noindent\textbf{(1) Cross-Model Robustness.} Table~\ref{tab:video_s3} reports the S@3 for recovering the seed words using different video generators and recognizers. Among the generators, videos produced by Veo paired with Gemini~3.1 Pro achieve the highest success rate at 95\%. Sora and Seedance also maintain relatively stable performance. Overall, Gemini~3.1 Pro demonstrates the strongest robustness as a recognizer, while the GPT series models show relatively lower success rates.

\begin{table}[t]
\centering
\caption{S@3 for recovering the seed words using different video generators and recognizers.}
\label{tab:video_s3}
\renewcommand{\arraystretch}{1.3}
\resizebox{\columnwidth}{!}{
\begin{tabular}{ccccc c}
\toprule
\multirow{2}{*}{\textbf{Video Generator}}
 & \multicolumn{5}{c}{\textbf{Video Recognizer}} \\
\cmidrule(lr){2-6}
 & \shortstack[c]{Gemini\\[-1pt]3.1 Flash}
 & \shortstack[c]{Gemini\\[-1pt]3.1 Pro}
 & \shortstack[c]{GPT\\[-1pt]5.3}
 & \shortstack[c]{GPT\\[-1pt]5.4(Std)}
 & \shortstack[c]{GPT\\[-1pt]5.4(Ext)} \\
\midrule
Veo      & 89\% & \textbf{95\%} & 73\% & 82\% & 84\% \\
Sora     & 82\% & 84\% & 70\% & 77\% & 78\% \\
Seedance & 83\% & 87\% & 68\% & 76\% & 80\% \\
\bottomrule
\end{tabular}
}
\end{table}

\noindent\textbf{(2) Time Consumption.} Table~\ref{tab:video_time} reports the time required for successful extractions, measured in seconds. The choice of video generator heavily influences the total time. Seedance requires significantly more time, often pushing the extraction process above 500 seconds, whereas Veo is the most efficient across all recognizers. Among the recognizers, GPT~5.3 responds the fastest as an instant model without an explicit thinking phase, but this speed is accompanied by lower accuracy. Furthermore, although the GPT~5.4 Thinking variants consume more time for their extended thinking process, this mode does not significantly improve the extraction accuracy. This demonstrates that the strong thinking ability of GPT does not offer as much advantage for this task as the strong multimodal understanding of Gemini.

\begin{table}[t]
\centering
\caption{Time consumption for successful recovery of the seed words using different video generators and recognizers.}
\label{tab:video_time}
\renewcommand{\arraystretch}{1.3}
\resizebox{\columnwidth}{!}{
\begin{tabular}{ccccc c}
\toprule
\multirow{2}{*}{\textbf{Video Generator}}
 & \multicolumn{5}{c}{\textbf{Video Recognizer}} \\
\cmidrule(lr){2-6}
 & \shortstack[c]{Gemini\\[-1pt]3.1 Flash}
 & \shortstack[c]{Gemini\\[-1pt]3.1 Pro}
 & \shortstack[c]{GPT\\[-1pt]5.3}
 & \shortstack[c]{GPT\\[-1pt]5.4(Std)}
 & \shortstack[c]{GPT\\[-1pt]5.4(Ext)} \\
\midrule
Veo      & 139 & 173 & \textbf{133} & 207 & 212 \\
Sora     & 205 & 261 & 197 & 294 & 298 \\
Seedance & 541 & 583 & 516 & 619 & 633 \\
\bottomrule
\end{tabular}
}
\end{table}

\noindent\textbf{(3) Discussion.}
Overall, these results support the feasibility of extending the method from images to video.
At the same time, compared with the image setting, video currently brings higher cost and lower stability.
We therefore regard it as a promising extension, while keeping the image setting as the main configuration.

%% file: sections/appendix.tex
\section{Case Studies}
\label{app:case_study}

In this section, we provide examples of the images generated by three different multimodal large language models (Gemini-3 Pro, Qwen Image Max, and GPT4o). For each case, we display the image and the corresponding stego caption generated by GPT5.1.

% ---- Case 1: Gemini ----
\subsection{Gemini-3 Pro}

\begin{figure}[!htbp]
    \centering
    \includegraphics[width=0.85\linewidth]{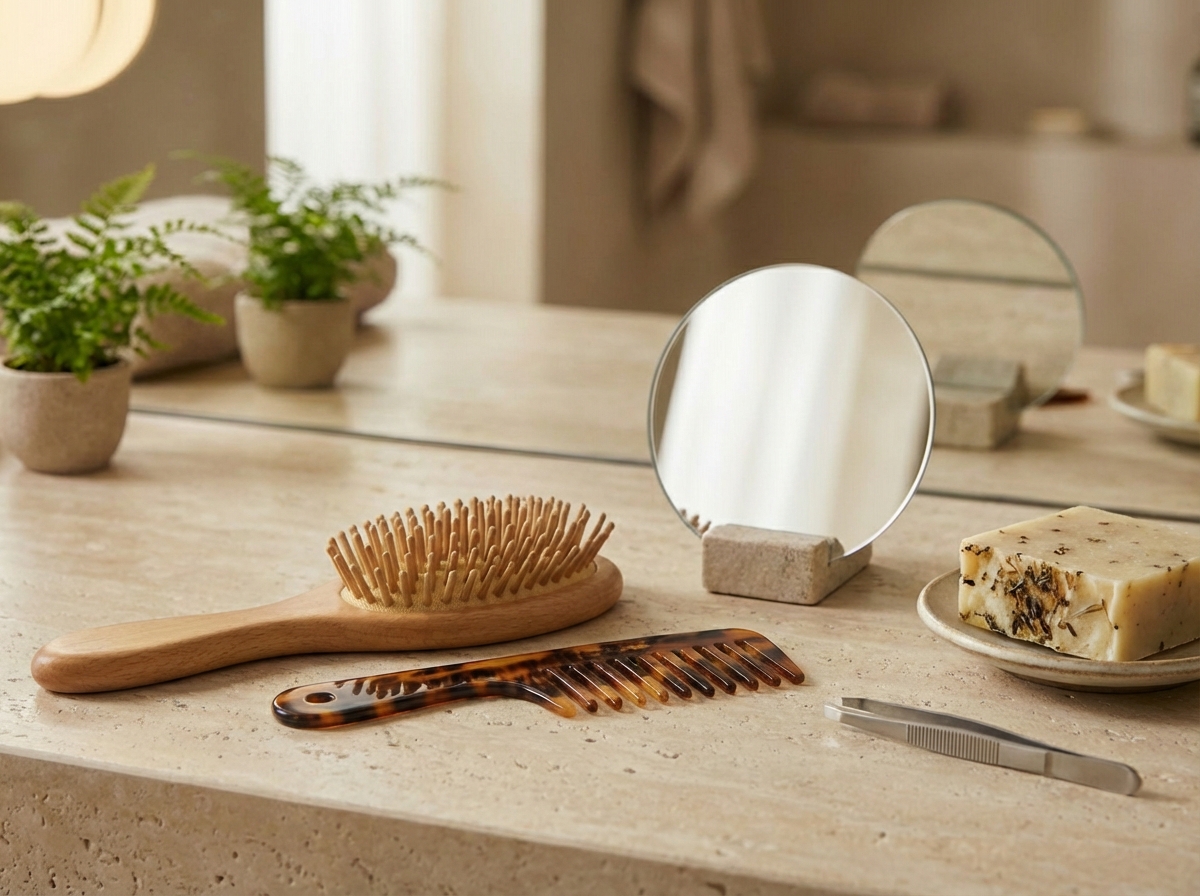}
    \caption{Image generated by Gemini-3 Pro.}
    \label{fig:case_gemini}
\end{figure}

\textbf{Stego Caption:} \textit{``Look at this wooden brush and patterned comb placed very carefully before a round mirror, sitting beside some textured soap and metallic tweezers.''}

% ---- Case 2: Qwen-Image-Max ----
\subsection{Qwen Image Max}

\begin{figure}[!htbp]
    \centering
    \includegraphics[width=0.85\linewidth]{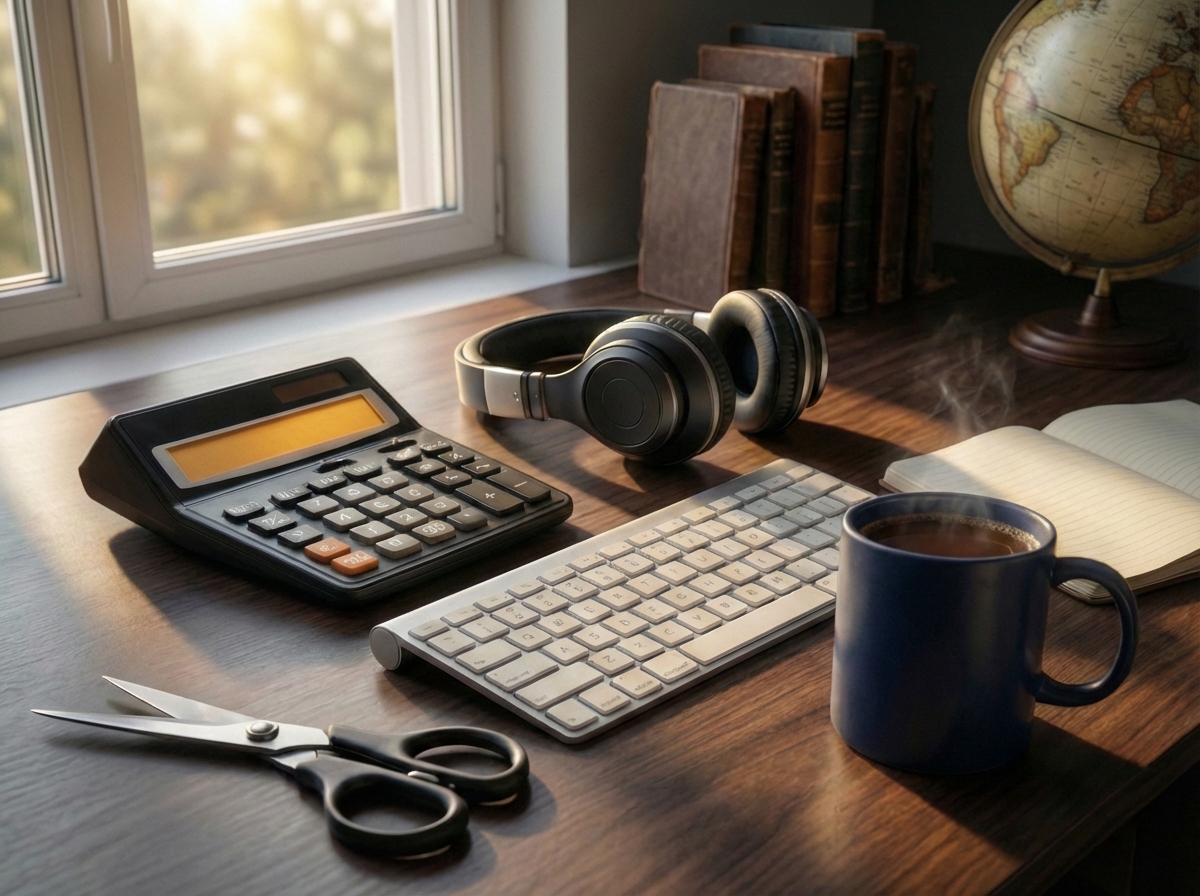}
    \caption{Image generated by Qwen Image Max.}
    \label{fig:case_qwen}
\end{figure}

\textbf{Stego Caption:} \textit{``See this dark calculator, these comfortable headphones, that wide keyboard, a hot mug, and sharp scissors. They sit peacefully, waiting for you to begin.''}

% ---- Case 3: GPT-4o ----
\subsection{GPT4o}

\begin{figure}[!htbp]
    \centering
    \includegraphics[width=0.85\linewidth]{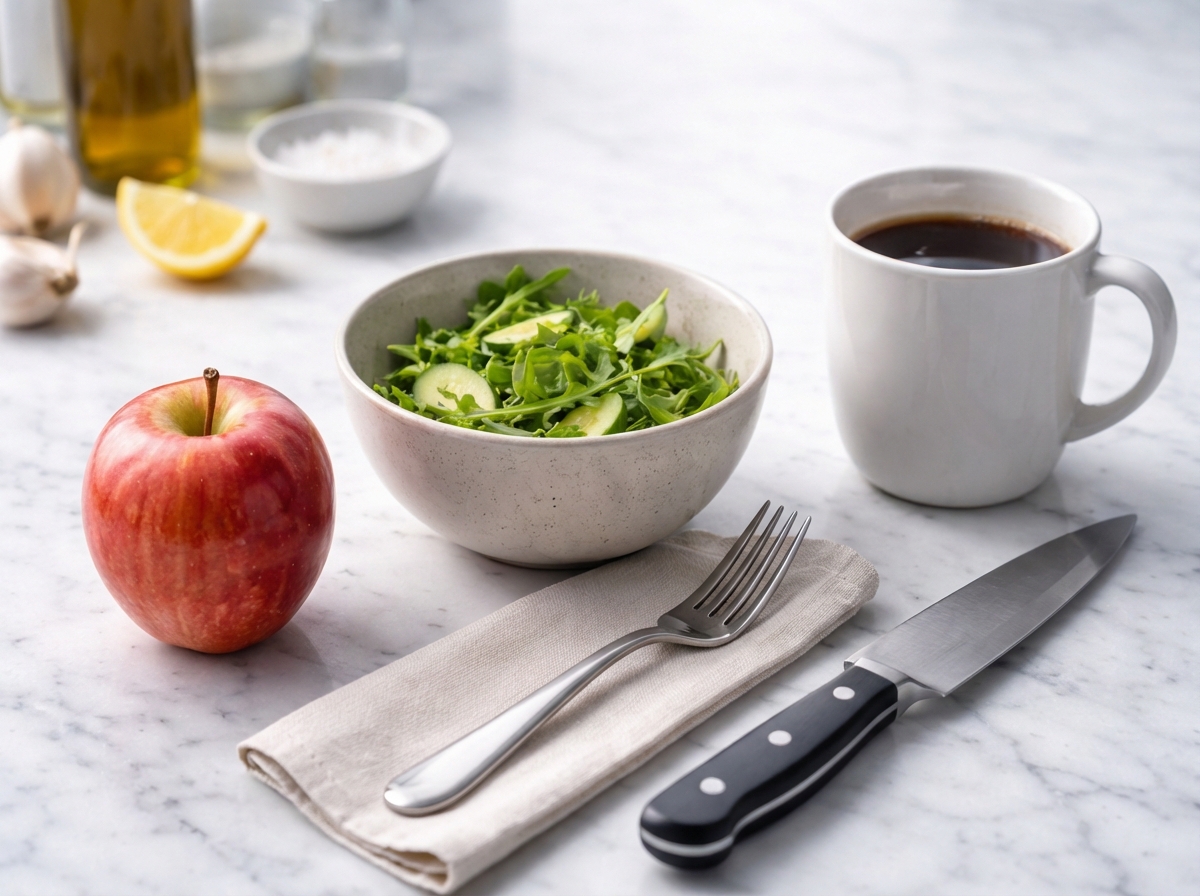}
    \caption{Image generated by GPT4o.}
    \label{fig:case_gpt}
\end{figure}

\textbf{Stego Caption:} \textit{``A fresh red apple sits near the full bowl, while a clean fork and a sharp knife lie ready next to the warm mug.''}

%===========================================================
\section{Details of Seeds Expansion}
\label{app:expansion_details}

This section gives a concrete and fully reproducible instantiation of the seeds expansion stage summarized in Section 3.2 of the main text. The main text intentionally presents this stage at a high level, whereas here we spell out the specific scoring and ranking convention used in our experiments. In practice, Alice and Bob may agree on other deterministic implementations; the purpose of this section is to make one concrete realization explicit and reproducible. To maintain high semantic relevance and stable codebook quality, DyCo-Stega adopts a scoring mechanism that incorporates length-sensitive penalties, field priorities, and frequency smoothing. We detail the normalization pipeline, the three expansion routes, the scoring rules, and the deterministic selection and ranking steps that produce the final semantic pool $V_{sem}$.

\subsection{Normalization and Dictionary Preprocessing}

We use a normalization function $\Norm(\cdot)$ throughout the entire procedure. For any token or headword, $\Norm(\cdot)$ applies three operations in order: lowercasing, removal of non-alphabetic characters, and WordNet lemmatization in noun form. All matching decisions in this appendix are performed only after this normalization step.

For each dictionary entry, we process the definition, synonym, and example fields in the same way. We first tokenize the field text, then normalize each token with $\Norm(\cdot)$, and finally remove stopwords and single-character tokens. The resulting normalized token sets are denoted by $\Def(\cdot)$, $\Syn(\cdot)$, and $\Ex(\cdot)$, respectively.

Two implementation details are fixed to avoid ambiguity. First, seed words are always retained even if they would otherwise be removed as stopwords. Second, if a seed word is missing from the public dictionary $\mathcal{D}$, we still insert it as a valid entry so that both parties reconstruct the same candidate universe; in that case, its dictionary frequency is set to $0$.

\subsection{Candidate Generation from the Seed Set}

Let $\mathcal{K}_{ord}=\{k_1,...,k_K\}$ be the ordered seed word list recovered from the image. The candidate universe is built by applying three deterministic expansion routes around this seed set. A word enters the candidate universe as long as it is triggered by at least one seed under at least one matching route. After all candidates are collected, duplicates are removed after normalization.

\paragraph{Reverse-definition matching.}
A candidate word $w$ is linked to a seed word $k$ by reverse-definition evidence if the normalized seed $\Norm(k)$ appears in the normalized definition-token set of $w$, namely $\Def(w)$. Intuitively, this captures words whose dictionary definitions explicitly refer back to the seed.

This route uses a length-sensitive weight $\lambda_{rev}(w)$. The base value is inversely scaled by the definition length so that very long definitions do not receive excessive credit:
$$\lambda_{rev}(w)=\frac{30.0}{\log_2(|\Def(w)|+4)}.$$

\paragraph{Forward-context matching.}
A candidate word $w$ is linked to a seed word $k$ by forward-context evidence if the normalized candidate $\Norm(w)$ appears in at least one of the three normalized token sets extracted from the dictionary entry of $k$: $\Syn(k)$, $\Def(k)$, or $\Ex(k)$. In other words, the seed is used as a local context source, and any candidate mentioned in that context becomes eligible for support.

The weight of this route, $\lambda_{fwd}(w,k)$, depends on which field provides the match. We use the fixed precedence
$$\Syn(k)\ \succ\ \Def(k)\ \succ\ \Ex(k),$$
so if the same candidate appears in multiple fields, only the highest-priority field determines the weight. Concretely, a synonym match receives a weight of $15.0$, a definition match receives a weight of $12.0$, and an example match receives a weight of $2.5$.

\paragraph{Prefix matching.}
We use a lightweight deterministic prefix rule for morphological proximity. Let $\widehat{w}=\Norm(w)$ and $\widehat{k}=\Norm(k)$. The match fires when $\widehat{w}$ starts with $\widehat{k}$ and is strictly longer than $\widehat{k}$. This ensures that only non-trivial extensions of the seed are counted. The corresponding weight is fixed at:
$$\lambda_{pre} = 4.0.$$

\subsection{Seed-wise Scoring and Aggregation}

For each candidate word $w$ and each seed word $k \in \mathcal{K}_{ord}$, we record whether the three routes fire. We denote the resulting binary evidence vector by
$$\mathbf{e}(w,k)=\bigl(e^{\mathrm{rev}}_{w,k},\,e^{\mathrm{fwd}}_{w,k},\,e^{\mathrm{pre}}_{w,k}\bigr)^\top.$$
Here, each indicator is either $1$ or $0$, depending on whether the corresponding match is present.

The seed-wise score is computed by summing the weights of all active routes for the pair $(w,k)$. In addition, we apply a deterministic mutual-confirmation boost when reverse-definition evidence is corroborated by at least one additional matching route for the same seed. More precisely, if reverse-definition evidence is present and either forward-context or prefix evidence is also present, then the entire seed-wise score for that pair is multiplied by $2.0$; otherwise, no boost is applied and the multiplier remains $1.0$.

Using this convention, the seed-wise score can be written compactly as
$$\mathrm{Score}(w,k)=\mu_{w,k}\Bigl(
\lambda_{rev}(w)e^{\mathrm{rev}}_{w,k}
+\lambda_{fwd}(w,k)e^{\mathrm{fwd}}_{w,k}
+\lambda_{pre} e^{\mathrm{pre}}_{w,k}
\Bigr),$$
where $\mu_{w,k}\in\{1.0,2.0\}$ follows the rule above, and $\lambda_{fwd}(w,k)$ is determined by the field-priority rule described in the previous subsection.

We then aggregate support over all seed words:
$$S(w)=\sum_{k\in \mathcal{K}_{ord}} \mathrm{Score}(w,k).$$
At this point, $S(w)$ reflects how strongly the candidate is supported by the entire seed set, but it still does not distinguish between a candidate supported repeatedly by one seed and a candidate supported by several different seeds. We therefore introduce two global modifiers.

\paragraph{Hub reward.}
Let $N_w$ be the number of distinct seeds in $\mathcal{K}_{ord}$ that provide any non-zero evidence for $w$. This counts seed coverage rather than total match count. We reward broader support by
$$\rho(w)=1.0+1.5\,(N_w-1).$$
Thus, a candidate supported by multiple different seeds receives a larger final score than one supported by only a single seed, all else being equal.

\paragraph{Frequency smoother.}
Let $\Freq(w)$ denote the dictionary frequency of the normalized headword $w$. We use a monotone smoothing factor
$$\varphi(w)=1.0+0.8\log_{10}(\Freq(w)+2),$$
which mildly favors more common words without allowing raw frequency to dominate the semantic evidence.

\paragraph{Final relevance score.}
The final score used to rank candidates is
$$\mathcal{F}(w)=S(w)\cdot \rho(w)\cdot \varphi(w).$$

\subsection{Selection and Ranking}

After computing $\mathcal{F}(w)$ for all candidates, we construct the final semantic pool $V_{sem}$ following the same overall logic described in the main text.

First, we exclude the original seed words $\mathcal{K}_{ord}$ from the candidate universe to avoid duplication. We then select the $\alpha - K$ highest-scoring non-seed candidates based on their final relevance score $\mathcal{F}(w)$. If several words receive exactly the same score, ties are broken alphabetically in ascending order. This selected candidate set is denoted as $V_{cand}$.

To ensure a stable and linguistically natural internal order for the dynamic codebook, we then independently sort the words within $V_{cand}$ in descending order according to their original dictionary frequencies.

Finally, we construct the semantic pool $V_{sem}$ by placing the ordered seed word list $\mathcal{K}_{ord}$ in the first $K$ positions, immediately followed by the sorted candidate set $V_{cand}$:
$$V_{sem} = \mathcal{K}_{ord} \,\|\, V_{cand} = (w_1,\ldots,w_\alpha).$$

This deterministic procedure ensures that the most visually central words (the seeds) occupy the front portion of the pool, while the remaining high-quality semantic expansions follow in a stable, frequency-based order. Because the entire pipeline depends only on the public dictionary $\mathcal{D}$ and the shared seed words $\mathcal{K}_{ord}$, both Alice and Bob will reconstruct exactly the same semantic pool $V_{sem}$ without requiring any local model weights or embeddings.

%===========================================================
\section{Details of Seed-Based Ordering}
\label{app:seeded_permutation}

This section gives one concrete implementation example of the seed-based ordering step in Section 3.2. In practice, any deterministic seed-driven ordering rule jointly agreed by the communicating parties may be used. Here we adopt an HMAC-SHA256-based sorting construction because it is simple, deterministic, and easy to reproduce.

\subsection*{Algorithm}

The core idea is to assign a pseudorandom hash value to each index using the session seed $\xi$, and then sort the indices based on these hashes to achieve a deterministic shuffle. Given the session seed $\xi$ and the target codebook size $\alpha$, the permutation $\pi_{\xi}$ is generated through the following steps:

\begin{enumerate}
    \item \textbf{Tag Generation:} For each index $j \in \{1, \dots, \alpha\}$, compute a 256-bit pseudorandom tag using the HMAC-SHA256 algorithm keyed by the session seed $\xi$:
    $$v_j = \mathrm{HMAC\mbox{-}SHA256}(\xi, j).$$
    
    \item \textbf{Sorting:} Rank the indices $j$ in ascending order according to their corresponding tag values $v_j$.
    
    \item \textbf{Ordering Formation:} The resulting sorted sequence of indices forms the permutation $\pi_{\xi}$. For instance, if the sorted tags correspond to the original indices $(3, 1, 4, 2)$, the new ordering places the 3rd item first, the 1st item second, the 4th item third, and the 2nd item last.
\end{enumerate}

Thus, for a fixed session seed $\xi$, both the sender and the receiver deterministically derive the same shuffled permutation without requiring any additional communication.

%===========================================================
\section{Details of Repeated Sampling}
Since $\tilde p_j N$ is generally not an integer, a deterministic discretization rule is required. For each word $\tilde w_j$ in the semantic pool $V_{sem}$, we first assign the integer base length
\[
a_j = \big\lfloor \tilde p_j N \big\rfloor,
\]
and define the corresponding fractional remainder
\[
r_j = \tilde p_j N - a_j = \big\{ \tilde p_j N \big\},
\]
where $\lfloor \cdot \rfloor$ denotes the floor operator and $\{\cdot\}$ is the fractional part.
\\
Let the total base allocation be
\[
\mathrm{Base} = \sum_{j=1}^{\alpha} a_j
= \sum_{j=1}^{\alpha} \big\lfloor \tilde p_j N \big\rfloor.
\]
The remaining unassigned integer budget is then
\[
R = N - \mathrm{Base}.
\]
Because $\sum_{j=1}^{\alpha} \tilde p_j = 1$, we have
\[
R
= N - \sum_{j=1}^{\alpha} \big\lfloor \tilde p_j N \big\rfloor
= \sum_{j=1}^{\alpha} \big\{ \tilde p_j N \big\},
\]
and therefore $R$ is an integer satisfying
\[
0 \le R < \alpha.
\]

We adopt a deterministic prefix-remainder allocation rule: the first $R$ words in the reordered sequence each receive one additional integer unit, while the remaining words receive none. Formally, let
\[
\delta_j =
\begin{cases}
1, & j \le R,\\
0, & j > R.
\end{cases}
\]
The final interval length assigned to $\tilde w_j$ is
\[
L_j = \big\lfloor \tilde p_j N \big\rfloor + \delta_j.
\]
By construction,
\[
\sum_{j=1}^{\alpha} L_j
= \sum_{j=1}^{\alpha} \big\lfloor \tilde p_j N \big\rfloor + \sum_{j=1}^{\alpha} \delta_j
= \mathrm{Base} + R
= N.
\]
Hence, the discretized interval lengths exactly partition the whole integer range $[0,N)$.

After the interval lengths are fixed, the interval boundaries are defined cumulatively by
\[
B_0 = 0, \qquad
B_j = \sum_{t=1}^{j} L_t, \quad j = 1,2,\dots,\alpha.
\]
The interval associated with $\tilde w_j$ is then
\[
\Omega_j = [B_{j-1}, B_j).
\]
Therefore,
\[
|\Omega_j| = L_j,
\]
the intervals are pairwise disjoint, and
\[
\bigcup_{j=1}^{\alpha} \Omega_j = [0,N).
\]
Consequently, every integer index $\mathrm{idx} \in [0,N-1]$ falls into exactly one interval $\Omega_j$.

%===========================================================
\section{Details of XOR Masking}

This section specifies the bit-length convention used in the XOR masking step, so that the private key has a fixed and unambiguous length.

Suppose that $\tau$ codewords are embedded in one stego caption. For the $i$-th embedded codeword, let the selected interval be
\[
\Omega_{j_i} = [B_{j_i-1}, B_{j_i}),
\]
and let the corresponding integer index be $\mathrm{idx}_i$.
The within-interval offset is defined as
\[
o_i = \mathrm{idx}_i - B_{j_i-1}.
\]
Since
\[
0 \le o_i < L_{j_i} \le N = 2^b,
\]
each offset $o_i$ admits a unique fixed-width $b$-bit binary representation.

We denote by $\mathrm{Int2Bin}(x)$ the standard binary representation of a nonnegative integer $x$, and by $|\mathrm{Int2Bin}(x)|$ its bit length.
For offset encoding, we use the canonical fixed-width $b$-bit representation
\[
\mathrm{Bin}_b(x)
=
0^{\,b-|\mathrm{Int2Bin}(x)|} \,\|\, \mathrm{Int2Bin}(x),
\qquad 0 \le x < 2^b,
\]
where $\|$ denotes bit-string concatenation.
That is, leading zeros are added when necessary so that the encoded offset always has exactly $b$ bits.

For the session seed, we derive a $b$-bit masking string from $\xi$ by the deterministic rule
\[
\mathrm{Seed}_b(\xi)
=
\begin{cases}
\mathrm{prefix}_b\!\big(\mathrm{Int2Bin}(\xi)\big), & |\mathrm{Int2Bin}(\xi)| \ge b,\\[4pt]
0^{\,b-|\mathrm{Int2Bin}(\xi)|}\,\|\,\mathrm{Int2Bin}(\xi), & |\mathrm{Int2Bin}(\xi)| < b,
\end{cases}
\]

where $\mathrm{prefix}_b(\cdot)$ returns the first $b$ bits of the input bit string.
Thus, the masking string used in each XOR operation also has exactly $b$ bits.

The private key fragment corresponding to the $i$-th codeword is then defined as
\[
S_i = \mathrm{Bin}_b(o_i) \oplus \mathrm{Seed}_b(\xi),
\]
where $\oplus$ denotes bitwise XOR.
Because both operands are $b$-bit strings, each fragment $S_i$ also has length exactly $b$.

The complete private key is formed by concatenating the $\tau$ fragments:
\[
S = S_1 \,\|\, S_2 \,\|\, \cdots \,\|\, S_\tau.
\]
Therefore, the total length of the private key is fixed as
\[
|S| = \tau b.
\]

During extraction, the receiver splits $S$ into $\tau$ consecutive $b$-bit blocks and recovers each offset by
\[
\mathrm{Bin}_b(o_i) = S_i \oplus \mathrm{Seed}_b(\xi),
\]
followed by
\[
o_i = \mathrm{Bin2Int}\!\big(\mathrm{Bin}_b(o_i)\big).
\]
Hence, each embedded codeword is associated with one fixed-length private-key fragment, and the complete private key is an unambiguous concatenation of $\tau$ such fragments.

%===========================================================
\section{Algorithm of Stego Caption Generation}

After determining the codeword, the sender uses a multimodal LLM to generate a stego caption for the image. For reliable extraction, the generated caption must satisfy three conditions: it must contain the selected codeword $w$, it must contain the anchor sequence $\AncSeq$, and it must not contain any other words from the semantic pool $V_{\sem}$. We use a verification function $\textsc{Verify}(C; w, \AncSeq, V_{\sem})$ to check these conditions.

First, an embedding prompt guides the multimodal LLM to generate a candidate stego caption $C$. If $C$ fails verification, a caption feedback prompt asks the LLM to identify the errors and produce an update prompt for revision. This loop continues until the caption passes verification. The accepted stego caption is then posted on the OSN.

\begin{algorithm}[t]
\caption{Stego caption generation based on reject sampling}
\label{alg:stego-caption}
\begin{algorithmic}[1]
\Require Image $I$, codeword $w$, anchor sequence $\AncSeq$, semantic pool $V_{\sem}$, embedding prompt template, caption feedback prompt template, multimodal LLM
\Ensure Stego caption $C$

\State $\mathit{accepted} \gets \mathrm{False}$
\State $P^{\emb} \gets$ construct embedding prompt using embedding prompt template, $w$, $\AncSeq$, and $V_{\sem}$
\State $C \gets$ generate a stego caption using multimodal LLM with image $I$ and $P^{\emb}$

\While{not $\mathit{accepted}$}
    \If{$\textsc{Verify}(C;\, w,\, \AncSeq,\, V_{\sem})$ = True}
        \State $\mathit{accepted} \gets \mathrm{True}$
    \Else
        \State $P^{\mathrm{fb}}_{C} \gets$ construct caption feedback prompt using caption feedback prompt template, $C$, $w$, $\AncSeq$, and $V_{\sem}$
        \State $P^{upd}_{C} \gets$ get update prompt using multimodal LLM with $C$, $w$, $\AncSeq$, $V_{\sem}$, and $P^{\mathrm{fb}}_{C}$
        \State $C \gets$ refine stego caption $C$ using multimodal LLM with $P^{upd}_{C}$
    \EndIf
\EndWhile

\If{$\mathit{accepted}$}
    \State \Return Stego caption $C$
\EndIf
\end{algorithmic}
\end{algorithm}
%===========================================================
\section{Mapping Rules for Session Parameters}
\label{app:private-convention}

This section provides one concrete implementation example of the one-time private agreement mentioned in Section 3.1 of the main text. In practice, the communicating parties may adopt any mutually agreed deterministic derivation rule. Here we present the specific bit-slicing convention used in our experiments. To achieve precise synchronization between the sender and receiver in a pure black-box, negotiation-free communication setting, DyCo-Stega uses the public image-caption pair and a pre-shared key $PSK \in \{0,1\}^{256}$. Through cryptographic hashing and bit-slicing, it deterministically derives the steganographic parameter set $\Phi=(\alpha,b,\sigma)$ and the session parameter $\Theta$. This procedure concretely instantiates the derivation function $g$ defined in Eq. (2) of the main text. The derivation proceeds as follows.

\noindent\textbf{1. Public Signal Extraction.}
The communicating parties first extract features from the publicly observable carriers. For visual features, the seed words are extracted from the public image $I$ via the multimodal LLM. These seed words are then sorted in descending order by their word frequency in the public dictionary $D$ to obtain the ordered seed word list $\mathcal{K}_{ord}$. For textual features, the anchor sequence $AncSeq$ is extracted from the public caption $C$.

\noindent\textbf{2. Cryptographic Hash Derivation.}
The high-entropy pre-shared key $PSK$, the ordered seed word list $\mathcal{K}_{ord}$, and the anchor sequence $AncSeq$ are serialized with a fixed mutually agreed byte-level convention, concatenated, and then fed into a secure hash function SHA-512:
\begin{equation}
H_{512}=\mathcal{H}\!\left(PSK \,\|\, \mathcal{K}_{ord} \,\|\, AncSeq\right).
\label{eq:app_hash_512}
\end{equation}
This outputs a 512-bit high-entropy bitstream $H_{512}$, which serves as the entropy source for all subsequent parameters. Note that Alice must specify $AncSeq$ before generating the stego caption.

\noindent\textbf{3. Deterministic Bit-Slicing and Parameter Parsing.}
Specific contiguous bit segments of $H_{512}$ are parsed into the session parameters as follows:
\begin{itemize}
    \item \textbf{Codebook size $\alpha$.} Extract bits $1$--$6$ and convert them to an integer $v_{\alpha}\in[0,63]$.
    \[
    \alpha=v_{\alpha}+24,
    \]
    so that $\alpha\in[24,87]$.

    \item \textbf{Word bit capacity $b$.} Extract bits $7$--$12$ and convert them to an integer $v_b\in[0,63]$.
    \[
    b=v_b+24,
    \]
    so that $b\in[24,87]$.

    \item \textbf{Step multiplier $\gamma$.} Extract bits $13$--$16$ and convert them to an integer $v_m\in[0,15]$.
    \[
    \gamma=0.25\times(v_m+1),
    \]
    so that $\gamma\in[0.25,4.0]$ with a step size of $0.25$.

    \item \textbf{Probability shaping parameter $\sigma$.} Given the parsed multiplier and the number of codewords embedded in the stego caption, denoted as $\tau$,
    \[
    \sigma=\gamma\tau.
    \]

    \item \textbf{Session parameter $\Theta$.} Extract the last $256$ bits of $H_{512}$ (bits $257$--$512$) and use them directly as the unique parameter
    \[
    \Theta\in\{0,1\}^{256}
    \]
    for the current session.
\end{itemize}

\noindent\textbf{4. Session Seed Generation.}
After parsing the session parameters, both parties concatenate them to compute the integer session seed $\xi$ using a cryptographic hash function $\mathcal{H}(\cdot)$ and an integer mapping function $\mathit{Hash2Int}(\cdot)$, exactly as defined in Eq. (3) of the main text:
\begin{equation}
\xi=\mathit{Hash2Int}(\mathcal{H}(\alpha \,\|\, b \,\|\, \sigma \,\|\, \Theta)).
\label{eq:app_hash2int}
\end{equation}
This session seed $\xi$ determines the pseudorandom permutation $\pi_{\xi}$ and the interval partitioning mechanism within the Dynamic Codebook $C_D$.

%===========================================================
\section{Discussion of Recovery Cost}
\label{app:leakage_discussion}

We consider the following concrete scenario: if the private key $S$ is intercepted, how much uncertainty still remains for a warden who attempts to recover the hidden secret message without the pre-shared key $PSK$?

\subsection{Assumptions}

We adopt the following assumptions.

First, we assume a Kerckhoffs-style setting: the warden knows the public image, the public caption, the public dictionary $\mathcal{D}$, the seed expansion rule, the dynamic codebook construction process, and the concrete parameter-mapping convention described in Section~\ref{app:private-convention}. In other words, the secrecy is not assumed to come from hiding the algorithm itself.

Second, we assume that the warden intercepts the private key $S$, and $S$ is composed of $\tau$ consecutive segments, one for each embedded codeword, and each segment has length $b$. Therefore,
\begin{equation}
|S|=\tau b.
\label{eq:leak_length}
\end{equation}
Since the caption structure reveals the number of embedded codewords $\tau$, the warden can directly infer
\begin{equation}
b=\frac{|S|}{\tau}.
\label{eq:b_infer}
\end{equation}
Thus, in the present discussion, $b$ does not need to be brute-forced once $S$ is intercepted.

Third, we assume that the warden does \emph{not} know the pre-shared key $PSK$. Consequently, the warden cannot directly derive the correct session parameters $(\alpha,\sigma)$, $\Theta$, or the session seed $\xi$ from the public image, caption and the intercepted private key $S$.

Finally, we assume that the warden doesn't have a reliable validation oracle telling which candidate reconstruction is correct. Therefore, the quantities below should be understood as ambiguity counts and exhaustive search burdens under the stated assumptions.

\subsection{Ambiguity of Dynamic Codebooks}

Under the concrete implementation in Section~\ref{app:private-convention}, once $b$ has been inferred from Eq.~(\ref{eq:b_infer}), the remaining session parameters that still need to be identified are $\alpha$ and $\sigma$.

\paragraph{Parameter ambiguity.}
The codebook size $\alpha$ is parsed from a 6-bit segment, so it can take $64$ possible values:
\begin{equation}
\alpha \in \{24,25,\ldots,87\}.
\label{eq:alpha_range}
\end{equation}
The shaping multiplier is parsed from a 4-bit segment, which yields $16$ possible values. Since $\sigma=\gamma\tau$ and $\tau$ is observable, the warden must still consider $16$ possible values of $\sigma$:
\begin{equation}
\sigma \in \{0.25\tau,\,0.50\tau,\,\ldots,\,4.00\tau\}.
\label{eq:sigma_range}
\end{equation}

\paragraph{Ordering ambiguity.}
Even after fixing a candidate pair $(\alpha,\sigma)$, the warden still does not know the correct session seed $\xi$, which determines the seed-based ordering of the semantic pool and therefore the final layout of the dynamic codebook. At the codebook-layout level, we count this uncertainty by the number of possible permutations of the $\alpha$ words, namely,
\begin{equation}
\alpha !
\label{eq:perm_alpha}
\end{equation}
If the warden wants to account for \emph{all} admissible values of $\alpha$, then the total number of candidate branches is
\begin{equation}
M_D
=
\sum_{\alpha=24}^{87}\sum_{u=1}^{16}\alpha!
=
16\sum_{\alpha=24}^{87}\alpha!.
\label{eq:mstruct_exact}
\end{equation}

\paragraph{A conservative lower bound.}
If one only wants a simple lower bound, one may use the fact that $\alpha!\ge 24!$ for every $\alpha\in\{24,\ldots,87\}$. Therefore,
\begin{equation}
M_D
=
16\sum_{\alpha=24}^{87}\alpha!
\;\ge\;
64\times 16\times 24!.
\label{eq:mstruct_lower}
\end{equation}
The right-hand side is \emph{not} the exact total number of candidates; rather, it is a conservative lower bound obtained by replacing every $\alpha!$ term with the smallest admissible factorial $24!$.

Numerically,
\begin{equation}
64\times16\times24!\approx 2^{89.04}.
\label{eq:mstruct_lower_log}
\end{equation}

\subsection{Ambiguity of Selected Intervals}

Even if the warden fixes a candidate structural branch and can identify the $\tau$ observed codewords in the caption, this still does not uniquely determine the embedded message bits. The reason is that each observed codeword only determines the corresponding interval in the dynamic codebook, while the exact offset inside that interval remains unknown.

Let the dynamic codebook under a candidate branch be
\[
C_D=\{(\widetilde{w}_j,\Omega_j)\}_{j=1}^{\alpha},
\]
where the length of interval $\Omega_j$ is denoted by
\[
L_j = |\Omega_j|.
\]
Since the codebook spans the integer range $[0,N)$ with
\begin{equation}
N=2^b,
\label{eq:N_2b}
\end{equation}
the interval lengths satisfy
\begin{equation}
\sum_{j=1}^{\alpha} L_j = N.
\label{eq:sum_Lj}
\end{equation}

For the $t$-th embedded position, let $J_t\in\{1,\dots,\alpha\}$ denote the index of the interval selected by the hidden message. Then the probability that the $t$-th position falls into interval $\Omega_j$ is
\begin{equation}
\Pr(J_t=j)=\frac{L_j}{N}.
\label{eq:Jt_prob}
\end{equation}

We define
\begin{equation}
M_I=\prod_{t=1}^{\tau} L_{J_t},
\label{eq:MX_def}
\end{equation}
which represents the number of candidate offset combinations consistent with the $\tau$ observed codewords under the current branch.

Assuming independence across positions, we obtain
\begin{equation}
\begin{aligned}
\mathbb{E}[M_I] 
&= \mathbb{E}\!\left[\prod_{t=1}^{\tau} L_{J_t}\right] = \prod_{t=1}^{\tau}\mathbb{E}[L_{J_t}] \\
&= \left(\sum_{j=1}^{\alpha}\Pr(J_t=j)L_j\right)^{\tau} = \left(\frac{1}{N}\sum_{j=1}^{\alpha}L_j^2\right)^{\tau}.
\end{aligned}
\label{eq:MX_exact}
\end{equation}

Considering that
\begin{equation}
L_j = N p_j,
\label{eq:Lj_approx}
\end{equation}
where $p_j$ is the probability assigned to the $j$-th codeword by the shaping rule. Substituting Eq.~(\ref{eq:Lj_approx}) into Eq.~(\ref{eq:MX_exact}) yields
\begin{equation}
\mathbb{E}[M_I]
=
\left(
\frac{1}{N}\sum_{j=1}^{\alpha}(N p_j)^2
\right)^{\tau}
=
\left(
N\sum_{j=1}^{\alpha}p_j^2
\right)^{\tau}
=
\left(
2^b\sum_{j=1}^{\alpha}p_j^2
\right)^{\tau}.
\label{eq:MX_pj}
\end{equation}

Under the half-Gaussian shaping rule,
\begin{equation}
p_j
=
\frac{
\exp\!\left(
-\frac{(j-1)^2}{2\sigma^2}
\right)
}{
\sum_{i=1}^{\alpha}
\exp\!\left(
-\frac{(i-1)^2}{2\sigma^2}
\right)
},
\qquad j=1,\dots,\alpha.
\label{eq:half_gaussian_pj}
\end{equation}
Hence,
\begin{equation}
\sum_{j=1}^{\alpha} p_j^2
=
\frac{
\sum_{j=1}^{\alpha}
\exp\!\left(
-\frac{(j-1)^2}{\sigma^2}
\right)
}{
\left(
\sum_{i=1}^{\alpha}
\exp\!\left(
-\frac{(i-1)^2}{2\sigma^2}
\right)
\right)^2
}.
\label{eq:sum_pj_square}
\end{equation}
Substituting Eq.~(\ref{eq:sum_pj_square}) into Eq.~(\ref{eq:MX_pj}), we obtain
\begin{equation}
\mathbb{E}[M_X]
\approx
\left(
2^b\cdot
\frac{
\sum_{j=1}^{\alpha}
\exp\!\left(
-\frac{(j-1)^2}{\sigma^2}
\right)
}{
\left(
\sum_{i=1}^{\alpha}
\exp\!\left(
-\frac{(i-1)^2}{2\sigma^2}
\right)
\right)^2
}
\right)^{\tau}.
\label{eq:MX_halfgaussian}
\end{equation}

Equation~(\ref{eq:MX_halfgaussian}) characterizes the representative scale of the offset search space induced by the half-Gaussian shaping rule. Under the default parameter setting used in the paper,
\[
(\alpha,b,\sigma)=(24,28,2.5), \qquad \tau=5,
\]
We obtain
\begin{equation}
\mathbb{E}[M_X]\approx 2^{128.6}.
\label{eq:MX_numeric}
\end{equation}
This quantity is used here to characterize the representative order of magnitude of the offset search space under the current parameter setting.

\subsection{Combined Interpretation}

Combining the conservative structural lower bound in Eq.~(\ref{eq:mstruct_lower_log}), we obtain the following representative estimate:
\begin{equation}
M
=
M_D · M_I
\approx 2^{217.6}.
\label{eq:mrep}
\end{equation}

Overall, intercepting the private key $S$ does not reduce the recovery problem to a simple inversion task. Even under this scenario, the warden must still cope with substantial ambiguity in reconstructing the dynamic codebook, as well as a large search space within the selected intervals.

%===========================================================
\section{Ethical Considerations and Broader Impact}

This work studies multimodal generative steganography in a black-box, API-only setting. As such, it is inherently dual-use. On the one hand, our study provides a concrete framework for analyzing the security, deployability, and forensic properties of steganographic communication when modern multimodal models are accessed only through public interfaces. We believe that making these risks explicit can help the community better understand emerging threat models in multimodal systems and motivate stronger defensive research, including multimodal steganalysis, platform-side auditing, and policy-aware robustness evaluation.

On the other hand, a system that reconstructs session-specific codebooks from public image--caption context and remains robust under cross-model transfer and OSN transmission could be misused to conceal malicious communication, evade moderation, or complicate forensic inspection. Our goal is not to facilitate such misuse, but to characterize a realistic capability boundary of black-box multimodal models and to encourage corresponding safeguards. For this reason, we do not advocate real-world deployment in safety-critical or adversarial settings, and we emphasize that future work should evaluate both attack and defense perspectives in parallel. In particular, stronger detection methods, auditing tools, and usage policies for multimodal generation platforms are needed as these systems become more capable and more widely accessible.

\newpage
\onecolumn

\section{Seed Words and Prompts We Used}
\label{app:scenes}
\label{app:prompts}

To ensure reliable extraction, we use predefined scene descriptions and seed words pool in Table \ref{tab:scene-presets} for image generation. By default, we sample exactly five seed words from this pool. Combining the scene description and seed words to construct the generation prompt $P^{\mathrm{gen}}$ ensures that the visual context remains controlled and the objects are semantically compatible with the scene.

Table~\ref{tab:prompts} lists the prompt templates used in DyCo-Stega. Operating under black-box MLLM access, DyCo-Stega relies heavily on instructional prompting and feedback mechanisms to enforce steganographic constraints. These prompts are divided into three primary functional groups: (1) the generation prompt template and extraction prompt that establish visual grounding and deterministic seed words recovery; (2) the embedding prompt template that guides the generation of the stego caption containing the required codeword and anchor sequences; and (3) feedback prompt templates that enable the models to self-correct visual or textual.

\begin{table}[h]
\centering
\small
\renewcommand{\arraystretch}{1.3}
\caption{Preset scene descriptions and their seed words pools used for image generation.}
\label{tab:scene-presets}
\begin{tabular}{@{} p{0.18\textwidth} p{0.48\textwidth} p{0.28\textwidth} @{}}
\toprule
\textbf{Scene} & \textbf{Scene description} & \textbf{Seed words pool} \\
\midrule

\texttt{office\_desk} &
An executive's organized workspace on a dark walnut wood desk. The lighting is warm morning sun filtered through a window, creating soft, long shadows. The aesthetic is ``Dark Academia'' meets modern minimalism. &
calculator, bread, headphones, keyboard, lamp, laptop, mouse, mug, notebook, pen, scissors, stapler \\
\addlinespace

\texttt{kitchen\_counter} &
A high-end culinary scene on a white Carrara marble countertop. The objects are dry and clean, lit by bright, cool-toned professional kitchen studio lights. The vibe is ``Food Magazine'' editorial photography. &
apple, banana, blender, bowl, fork, kettle, knife, lemon, mug, pan, spoon, toaster \\
\addlinespace

\texttt{bathroom\_vanity} &
A luxury spa-style vanity on a matte beige travertine stone surface. The lighting is soft and diffused (beauty lighting), creating a clean, serene, and airy atmosphere with soft-focus reflections. &
brush, candle, comb, hairdryer, lipstick, mirror, razor, soap, sponge, toothbrush, towel, tweezers \\

\bottomrule
\end{tabular}
\end{table}

\vspace{2em}

\renewcommand{\arraystretch}{1.0}
\setlength{\LTleft}{0pt}
\setlength{\LTright}{0pt}

\begin{longtable}{@{} p{\textwidth} @{}}
\caption{Prompt templates used in DyCo-Stega.}
\label{tab:prompts}\\

\toprule
\endfirsthead
\toprule
\endhead
\bottomrule
\endlastfoot

% ---------------------------------------------------------
% Prompt 1: Image Generation Prompt Template
% ---------------------------------------------------------
\textbf{Image Generation Prompt Template} \vspace{0.2em}\newline
\rule{\linewidth}{0.4pt} \vspace{0.5em}\newline
{\ttfamily\small
Create an ultra-realistic, high-resolution photograph of a natural, candid scene:

\{scene\_description\}

The image should look like a real photo taken by a skilled photographer, with believable perspective, natural depth of field, clean details, consistent lighting and shadows, and natural colors. Do not include CGI artifacts, glitches, distortions, text, UI elements, watermarks, logos, exaggerated effects, fantasy elements, or surreal content.

Keep the overall composition simple so that the viewer's main attention is directed to the specified foreground objects. If the scene description implies additional objects, include them only as small, soft, distant, or partially occluded background details.

Hard object constraints:\newline
- The image must clearly feature exactly five distinct main foreground objects, corresponding one-to-one to these words: \{items\_list\}.\newline
- For each word, include a single, clearly identifiable instance of that everyday object category.\newline
- Each target object must be fully visible, not cropped, not heavily overlapping with the others, large enough to be recognized, and in sharp focus.\newline
- Arrange the five target objects naturally in the foreground in a casual, slightly asymmetrical composition that fits the scene.

Additional realism constraints:\newline
- You may include minor supporting objects and background elements only to improve realism.\newline
- Do not add any other large, clear, visually separate foreground objects or any new object categories beyond \{items\_list\}.\newline
- Avoid prominent non-target items such as extra books, extra electronic devices, cups, plates, or tools.\newline
- Any additional objects must be smaller, farther away, less sharp, or partly hidden so that they are clearly less important than the five target objects.\newline
- Multi-part foods or similar items should appear as one coherent object rather than many separate pieces or slices.
} \\ 

% ---------------------------------------------------------
% Prompt 2: Extraction Prompt
% ---------------------------------------------------------
\midrule
\textbf{Extraction Prompt} \vspace{0.2em}\newline
\rule{\linewidth}{0.4pt} \vspace{0.5em}\newline
{\ttfamily\small
You are given one image that was created from a text prompt specifying exactly five simple English nouns, each naming one visible object. Your task is to recover those five nouns by looking for clearly visible, individual objects that are main subjects: large enough, in focus, not strongly hidden, and visually important. Treat multiple instances of the same kind as one category (many apples $\rightarrow$ apple; several chairs $\rightarrow$ chair), and ignore or treat as background walls, floor, ceiling, sky, grass, roads, mountains, sea, generic landscape, tiny decorations, textures, light patterns, reflections, text, logos, body parts, and objects that are mostly hidden or extremely small; do not invent objects that are not clearly present. First list all clearly visible, important object categories, then merge any categories that actually refer to the same kind of thing so the remaining categories are semantically distinct (for example book/books/novel $\rightarrow$ book; bread/slice/slices $\rightarrow$ bread; phone/cellphone/smartphone $\rightarrow$ phone); if there are still more than five candidates, keep the five most visually salient and semantically distinct ones and ignore the rest.

For each of the final five categories, normalize its name to one canonical noun and then output it. Use a single very common English noun in lowercase with no spaces, hyphens, numbers, adjectives (no colors, sizes, materials, styles), brands, or proper names; use the singular form even when several instances are visible, unless the plural is the usual dictionary form (glasses, scissors, sunglasses); never use words that describe only a part, portion, or arrangement of something else, such as slice, slices, piece, bit, half, bunch, pile, stack, pair, or set---always name the underlying object or food instead (bread not slices; cake not slice; headphones not pair); merge synonyms and specific variants into the most basic everyday term a non-expert would say (cellphone/mobile/smartphone $\rightarrow$ phone; notebook/computer/pc when clearly portable $\rightarrow$ laptop; novels/textbooks/storybooks $\rightarrow$ book; sofa/couch $\rightarrow$ sofa); prefer specific words over vague ones (apple not fruit; backpack not bag if clearly a backpack; guitar not instrument). Convert the five normalized categories to this form, sort them in alphabetical order, and respond with only a JSON array of five distinct strings, for example: ["apple","book","camera","laptop","mug"].
} \\ 

% ---------------------------------------------------------
% Prompt 3: Image Feedback Prompt Template
% ---------------------------------------------------------
\midrule
\textbf{Image Feedback Prompt Template} \vspace{0.2em}\newline
\rule{\linewidth}{0.4pt} \vspace{0.5em}\newline
{\ttfamily\small
You are given:\newline
- an existing image,\newline
- its original English prompt,\newline
- a list of five desired target nouns: \{items\_list\},\newline
- and a list of five nouns extracted from the current image: \{extracted\_list\}.

Task:\newline
Compare the target list and the extracted list, and convert any difference between them into object-level edit instructions for the image. Write exactly one concise English paragraph in imperative, edit-style language.

Requirements:\newline
- Do not rewrite or summarize the original prompt.\newline
- Do not describe the full scene again.\newline
- Assume that the base prompt and image remain unchanged; only specify how the objects should be modified so that a future noun-extraction step will return exactly the five target nouns and no other foreground nouns.\newline
- For every noun that appears in \{extracted\_list\} but not in \{items\_list\}, explicitly name that noun using exactly the given string and state that the corresponding object should be removed, strongly de-emphasized, or moved out of the main foreground.\newline
- For every noun that appears in \{items\_list\} but not in \{extracted\_list\}, explicitly name that noun using exactly the given string and state that the corresponding object should be added, enlarged, or made more prominent and clearly visible in the foreground.\newline
- Do not mention nouns that already appear in both lists.\newline
- Do not change the scene style, lighting, viewpoint, or background.\newline
- Do not introduce any new object categories beyond the target nouns.
} \\ 

% ---------------------------------------------------------
% Prompt 4: Embedding Prompt Template
% ---------------------------------------------------------
\midrule
\textbf{Embedding Prompt Template} \vspace{0.2em}\newline
\rule{\linewidth}{0.4pt} \vspace{0.5em}\newline
{\ttfamily\small
Write one natural-sounding English photo caption for the given image.

Scene context:\newline
\{scene\_description\}

Hard constraints:\newline
- The caption must include each required word exactly once: \{required\_words\}\newline
- The caption must contain the public anchor sequence in the specified order: \{anchor\_sequence\}\newline
- The caption must not include any forbidden word: \{forbidden\_words\}\newline
- The caption length must be between \{min\_len\} and \{max\_len\} words.\newline
- Word matching ignores case, simple plurals, and attached punctuation.

Writing guidelines:\newline
1. Write exactly one sentence.\newline
2. Start with the main visible subject; do not begin with ``This image shows'' or ``This image depicts''.\newline
3. Use simple, everyday English and plain grammar.\newline
4. Describe only what is visually supported by the image and the scene context.\newline
5. Prefer a natural photo-caption style, such as:\newline
   ``A/An + 3--6 visible objects + rest/sit/lie + on/along/beside + a surface + optional lighting phrase.''\newline
6. Place the required words naturally in the sentence.\newline
7. If a required word is only weakly grounded, attach it in a short final prepositional phrase such as ``with ...'' or ``near ...'', without adding unsupported claims.\newline
8. Avoid rare terms, awkward keyword stacking, and unsupported mood or marketing adjectives.

Output only the sentence.
} \\ 

% ---------------------------------------------------------
% Prompt 5: Caption Feedback Prompt Template
% ---------------------------------------------------------
\midrule
\textbf{Caption Feedback Prompt Template} \vspace{0.2em}\newline
\rule{\linewidth}{0.4pt} \vspace{0.5em}\newline
{\ttfamily\small
You are given:\newline
- the current caption C: \{current\_caption\}\newline
- the target ordered codeword sequence w* = \{required\_sequence\}\newline
- the public anchor sequence Anc-Seq = \{anchor\_sequence\}\newline
- the semantic keyword pool V\_sem = \{semantic\_pool\}

Task:\newline
Compare the current caption C against the protocol constraints below, and write minimal caption-edit instructions that will revise C into a valid stego caption.

Protocol constraints:\newline
1. C must contain the target codeword sequence w* in the same order.\newline
2. C must contain the anchor sequence Anc-Seq in the same order.\newline
3. C must not contain any keyword from V\_sem other than the target codewords required by w*.

Requirements:\newline
- Do NOT rewrite the full caption.\newline
- Do NOT explain your reasoning.\newline
- Do NOT describe the image again.\newline
- Output exactly one concise English paragraph in imperative, edit-style language.\newline
- Preserve as much of the current caption as possible, and keep the final caption as one fluent, natural-sounding photo caption.

Editing rules:\newline
- If any target codeword required by w* is missing, explicitly name it and instruct that it be added in a natural position.\newline
- If all target codewords are present but their order does not match w*, explicitly instruct that they be reordered to match the target sequence.\newline
- If any anchor token from Anc-Seq is missing, explicitly instruct that the missing anchor material be inserted.\newline
- If the anchor sequence appears but its order does not match Anc-Seq, explicitly instruct that it be reordered to match the target anchor sequence.\newline
- For every keyword from V\_sem that appears in C but is not part of the target realization of w*, explicitly name it and instruct that it be removed or paraphrased into a non-keyword alternative so that the surface word no longer matches any keyword in V\_sem.\newline
- If there are redundant extra occurrences of target codewords or anchor tokens beyond what is needed to realize the target ordered sequences, instruct that the extra occurrences be removed to avoid ambiguity.\newline
- Do not introduce any new keyword from V\_sem beyond the target codewords in w*.\newline
- Keep the edits minimal and local whenever possible.
} \\ 

\end{longtable}

%% file: ref.bib
@inproceedings{alayrac2022flamingo,
  author    = {Jean-Baptiste Alayrac and Jeff Donahue and Pauline Luc and Antoine Miech and Iain Barr and Yana Hasson and Karel Lenc and Arthur Mensch and Katherine Millican and Malcolm Reynolds and Roman Ring and Eliza Rutherford and Serkan Cabi and Tengda Han and Zhitao Gong and Sina Samangooei and Marianne Monteiro and Jacob L. Menick and Sebastian Borgeaud and Andy Brock and Aida Nematzadeh and Sahand Sharifzadeh and Miko{\l}aj Bi{\'n}kowski and Ricardo Barreira and Oriol Vinyals and Andrew Zisserman and Kar{\'e}n Simonyan},
  title     = {Flamingo: a Visual Language Model for Few-Shot Learning},
  booktitle = {Advances in Neural Information Processing Systems},
  volume    = {35},
  pages     = {23716--23736},
  publisher = {Curran Associates, Inc.},
  address   = {New Orleans, LA, USA},
  year      = {2022},
  doi       = {10.52202/068431-1723},
  url       = {https://proceedings.neurips.cc/paper_files/paper/2022/hash/960a172bc7fbf0177ccccbb411a7d800-Abstract-Conference.html}
}

@article{anderson1998limits,
  author  = {Ross J. Anderson and Fabien A. P. Petitcolas},
  title   = {On the Limits of Steganography},
  journal = {IEEE Journal on Selected Areas in Communications},
  volume  = {16},
  number  = {4},
  pages   = {474--481},
  year    = {1998},
  doi     = {10.1109/49.668971}
}

@misc{bai2025qwen25vl,
  author        = {Shuai Bai and Keqin Chen and Xuejing Liu and Jialin Wang and Wenbin Ge and Sibo Song and Kai Dang and Peng Wang and Shijie Wang and Jun Tang and Humen Zhong and Yuanzhi Zhu and Mingkun Yang and Zhaohai Li and Jianqiang Wan and Pengfei Wang and Wei Ding and Zheren Fu and Yiheng Xu and Jiabo Ye and Xi Zhang and Tianbao Xie and Zesen Cheng and Hang Zhang and Zhibo Yang and Haiyang Xu and Junyang Lin},
  title         = {{Qwen2.5-VL} Technical Report},
  year          = {2025},
  eprint        = {2502.13923},
  archiveprefix = {arXiv},
  primaryclass  = {cs.CV},
  url           = {https://arxiv.org/abs/2502.13923}
}

@article{chang2014practical,
  author  = {Ching-Yun Chang and Stephen Clark},
  title   = {Practical Linguistic Steganography Using Contextual Synonym Substitution and a Novel Vertex Coding Method},
  journal = {Computational Linguistics},
  volume  = {40},
  number  = {2},
  pages   = {403--448},
  year    = {2014},
  doi     = {10.1162/COLI_a_00176}
}

@inproceedings{chapman1997hiding,
  author    = {Mark Chapman and George I. Davida},
  title     = {Hiding the Hidden: A Software System for Concealing Ciphertext as Innocuous Text},
  booktitle = {Information and Communications Security ({ICICS} 1997)},
  series    = {Lecture Notes in Computer Science},
  volume    = {1334},
  pages     = {335--345},
  publisher = {Springer},
  address   = {Berlin, Heidelberg},
  year      = {1997},
  doi       = {10.1007/BFb0028489}
}

@book{cox2007digital,
  author    = {Ingemar J. Cox and Matthew L. Miller and Jeffrey A. Bloom and Jessica Fridrich and Ton Kalker},
  title     = {Digital Watermarking and Steganography},
  edition   = {2},
  publisher = {Morgan Kaufmann},
  address   = {Burlington, MA},
  year      = {2008},
  isbn      = {978-0-12-372585-1},
  doi       = {10.1016/B978-0-12-372585-1.X5001-3}
}

@inproceedings{dai2019near,
  author    = {Falcon Dai and Zheng Cai},
  title     = {Towards Near-Imperceptible Steganographic Text},
  booktitle = {Proceedings of the 57th Annual Meeting of the Association for Computational Linguistics},
  pages     = {4303--4308},
  publisher = {Association for Computational Linguistics},
  address   = {Florence, Italy},
  year      = {2019},
  doi       = {10.18653/v1/P19-1422}
}

@inproceedings{dai2023instructblip,
  author    = {Wenliang Dai and Junnan Li and Dongxu Li and Anthony Meng Huat Tiong and Junqi Zhao and Weisheng Wang and Boyang Li and Pascale Fung and Steven C. H. Hoi},
  title     = {{InstructBLIP}: Towards General-purpose Vision-Language Models with Instruction Tuning},
  booktitle = {Advances in Neural Information Processing Systems},
  volume    = {36},
  pages     = {49250--49267},
  publisher = {Curran Associates, Inc.},
  address   = {New Orleans, LA, USA},
  year      = {2023},
  doi       = {10.52202/075280-2142},
  url       = {https://proceedings.neurips.cc/paper_files/paper/2023/hash/9a6a435e75419a836fe47ab6793623e6-Abstract-Conference.html}
}

@inproceedings{ding2023discop,
  author    = {Jinyang Ding and Kejiang Chen and Yaofei Wang and Na Zhao and Weiming Zhang and Nenghai Yu},
  title     = {{Discop}: Provably Secure Steganography in Practice Based on ``Distribution Copies''},
  booktitle = {2023 IEEE Symposium on Security and Privacy},
  pages     = {2238--2255},
  publisher = {IEEE},
  address   = {San Francisco, CA, USA},
  year      = {2023},
  doi       = {10.1109/SP46215.2023.10179287}
}

@inproceedings{fang2017generating,
  author    = {Tina Fang and Martin Jaggi and Katerina Argyraki},
  title     = {Generating Steganographic Text with {LSTM}s},
  booktitle = {Proceedings of {ACL} 2017, Student Research Workshop},
  pages     = {100--106},
  publisher = {Association for Computational Linguistics},
  address   = {Vancouver, Canada},
  year      = {2017},
  url       = {https://aclanthology.org/P17-3017/}
}

@inproceedings{hessel2021clipscore,
  author    = {Hessel, Jack and Holtzman, Ari and Forbes, Maxwell and {Le Bras}, Ronan and Choi, Yejin},
  title     = {{CLIPScore}: A Reference-Free Evaluation Metric for Image Captioning},
  booktitle = {Proceedings of the 2021 Conference on Empirical Methods in Natural Language Processing},
  pages     = {7514--7528},
  publisher = {Association for Computational Linguistics},
  address   = {Online and Punta Cana, Dominican Republic},
  year      = {2021},
  doi       = {10.18653/v1/2021.emnlp-main.595}
}

@inproceedings{huang-etal-2026-od,
  author    = {Yu-Shin Huang and Peter Just and Hanyun Yin and Krishna Narayanan and Ruihong Huang and Chao Tian},
  title     = {{OD}-Stega: {LLM}-Based Relatively Secure Steganography via Optimized Distributions},
  booktitle = {Proceedings of the 19th Conference of the European Chapter of the Association for Computational Linguistics (Volume 1: Long Papers)},
  pages     = {827--851},
  publisher = {Association for Computational Linguistics},
  address   = {Rabat, Morocco},
  year      = {2026},
  doi       = {10.18653/v1/2026.eacl-long.36}
}

@inproceedings{kaptchuk2021meteor,
  author    = {Gabriel Kaptchuk and Tushar M. Jois and Matthew Green and Aviel D. Rubin},
  title     = {{METEOR}: Cryptographically Secure Steganography for Realistic Distributions},
  booktitle = {Proceedings of the 2021 ACM SIGSAC Conference on Computer and Communications Security},
  pages     = {1529--1548},
  publisher = {ACM},
  address   = {Virtual Event, Republic of Korea},
  year      = {2021},
  doi       = {10.1145/3460120.3484550}
}

@inproceedings{li2023blip2,
  author    = {Junnan Li and Dongxu Li and Silvio Savarese and Steven C. H. Hoi},
  title     = {{BLIP}-2: Bootstrapping Language-Image Pre-Training with Frozen Image Encoders and Large Language Models},
  booktitle = {Proceedings of the 40th International Conference on Machine Learning},
  series    = {Proceedings of Machine Learning Research},
  volume    = {202},
  pages     = {19730--19742},
  publisher = {PMLR},
  address   = {Honolulu, HI, USA},
  year      = {2023},
  url       = {https://proceedings.mlr.press/v202/li23q.html}
}

@inproceedings{lin-etal-2024-zero,
  author    = {Ke Lin and Yiyang Luo and Zijian Zhang and Luo, Ping},
  title     = {Zero-shot Generative Linguistic Steganography},
  booktitle = {Proceedings of the 2024 Conference of the North American Chapter of the Association for Computational Linguistics: Human Language Technologies (Volume 1: Long Papers)},
  pages     = {5168--5182},
  publisher = {Association for Computational Linguistics},
  address   = {Mexico City, Mexico},
  month     = jun,
  year      = {2024},
  doi       = {10.18653/v1/2024.naacl-long.289},
  url       = {https://aclanthology.org/2024.naacl-long.289/}
}

@misc{liu2019roberta,
  author        = {Yinhan Liu and Myle Ott and Naman Goyal and Jingfei Du and Mandar Joshi and Danqi Chen and Omer Levy and Mike Lewis and Luke Zettlemoyer and Veselin Stoyanov},
  title         = {{RoBERTa}: A Robustly Optimized {BERT} Pretraining Approach},
  year          = {2019},
  eprint        = {1907.11692},
  archiveprefix = {arXiv},
  primaryclass  = {cs.CL},
  url           = {https://arxiv.org/abs/1907.11692}
}

@inproceedings{liu2023visual,
  author    = {Haotian Liu and Chunyuan Li and Qingyang Wu and Yong Jae Lee},
  title     = {Visual Instruction Tuning},
  booktitle = {Advances in Neural Information Processing Systems},
  volume    = {36},
  pages     = {34892--34916},
  publisher = {Curran Associates, Inc.},
  address   = {New Orleans, LA, USA},
  year      = {2023},
  doi       = {10.52202/075280-1516},
  url       = {https://proceedings.neurips.cc/paper_files/paper/2023/hash/6dcf277ea32ce3288914faf369fe6de0-Abstract-Conference.html}
}

@article{peng2021realtime,
  author  = {Wanli Peng and Jinyu Zhang and Yiming Xue and Zhenghong Yang},
  title   = {Real-Time Text Steganalysis Based on Multi-Stage Transfer Learning},
  journal = {IEEE Signal Processing Letters},
  volume  = {28},
  pages   = {1510--1514},
  year    = {2021},
  doi     = {10.1109/LSP.2021.3097241}
}

@article{provos2003hide,
  author  = {Niels Provos and Peter Honeyman},
  title   = {Hide and Seek: An Introduction to Steganography},
  journal = {IEEE Security \& Privacy},
  volume  = {1},
  number  = {3},
  pages   = {32--44},
  year    = {2003},
  doi     = {10.1109/MSECP.2003.1203220}
}

@techreport{radford2019gpt2,
  author      = {Alec Radford and Jeffrey Wu and Rewon Child and David Luan and Dario Amodei and Ilya Sutskever},
  title       = {Language Models are Unsupervised Multitask Learners},
  institution = {OpenAI},
  year        = {2019},
  url         = {https://cdn.openai.com/better-language-models/language-models.pdf}
}

@inproceedings{radford2021clip,
  author    = {Alec Radford and Jong Wook Kim and Chris Hallacy and Aditya Ramesh and Gabriel Goh and Sandhini Agarwal and Girish Sastry and Amanda Askell and Pamela Mishkin and Jack Clark and Gretchen Krueger and Ilya Sutskever},
  title     = {Learning Transferable Visual Models From Natural Language Supervision},
  booktitle = {Proceedings of the 38th International Conference on Machine Learning},
  series    = {Proceedings of Machine Learning Research},
  volume    = {139},
  pages     = {8748--8763},
  publisher = {PMLR},
  address   = {Virtual},
  year      = {2021},
  url       = {https://proceedings.mlr.press/v139/radford21a.html}
}

@inproceedings{reimers2019sbert,
  author    = {Nils Reimers and Iryna Gurevych},
  title     = {Sentence-{BERT}: Sentence Embeddings Using Siamese {BERT}-Networks},
  booktitle = {Proceedings of the 2019 Conference on Empirical Methods in Natural Language Processing and the 9th International Joint Conference on Natural Language Processing},
  pages     = {3982--3992},
  publisher = {Association for Computational Linguistics},
  address   = {Hong Kong, China},
  year      = {2019},
  doi       = {10.18653/v1/D19-1410}
}

@inproceedings{schroeder2022perfectly,
  author    = {Schroeder de Witt, Christian and Samuel Sokota and J. Zico Kolter and Jakob Nicolaus Foerster and Martin Strohmeier},
  title     = {Perfectly Secure Steganography Using Minimum Entropy Coupling},
  booktitle = {The Eleventh International Conference on Learning Representations},
  year      = {2023},
  url       = {https://openreview.net/forum?id=HQ67mj5rJdR}
}

@inproceedings{ueoka2021masked,
  author    = {Honai Ueoka and Yugo Murawaki and Sadao Kurohashi},
  title     = {Frustratingly Easy Edit-Based Linguistic Steganography with a Masked Language Model},
  booktitle = {Proceedings of the 2021 Conference of the North American Chapter of the Association for Computational Linguistics: Human Language Technologies},
  pages     = {5486--5492},
  publisher = {Association for Computational Linguistics},
  address   = {Online},
  year      = {2021},
  doi       = {10.18653/v1/2021.naacl-main.433}
}

@inproceedings{wang2025sparsamp,
  author    = {Yaofei Wang and Gang Pei and Kejiang Chen and Jinyang Ding and Chao Pan and Weilong Pang and Donghui Hu and Weiming Zhang},
  title     = {{SparSamp}: Efficient Provably Secure Steganography Based on Sparse Sampling},
  booktitle = {34th {USENIX} Security Symposium ({USENIX} Security 25)},
  pages     = {6817--6835},
  publisher = {USENIX Association},
  address   = {Seattle, WA},
  month     = aug,
  year      = {2025},
  isbn      = {978-1-939133-52-6},
  url       = {https://www.usenix.org/conference/usenixsecurity25/presentation/wang-yaofei}
}

@article{wen2019cnn,
  author  = {Juan Wen and Xuejing Zhou and Ping Zhong and Yiming Xue},
  title   = {Convolutional Neural Network Based Text Steganalysis},
  journal = {IEEE Signal Processing Letters},
  volume  = {26},
  number  = {3},
  pages   = {460--464},
  year    = {2019},
  doi     = {10.1109/LSP.2019.2895286}
}

@article{wu2021gnn,
  author  = {Hanzhou Wu and Biao Yi and Feng Ding and Guorui Feng and Xinpeng Zhang},
  title   = {Linguistic Steganalysis With Graph Neural Networks},
  journal = {IEEE Signal Processing Letters},
  volume  = {28},
  pages   = {558--562},
  year    = {2021},
  doi     = {10.1109/LSP.2021.3062233}
}

@inproceedings{wu2024llmstega,
  author    = {Jiaxuan Wu and Zhengxian Wu and Yiming Xue and Juan Wen and Wanli Peng},
  title     = {Generative Text Steganography with Large Language Model},
  booktitle = {Proceedings of the 32nd ACM International Conference on Multimedia},
  pages     = {10345--10353},
  publisher = {ACM},
  address   = {Melbourne, VIC, Australia},
  year      = {2024},
  doi       = {10.1145/3664647.3680562}
}

@inproceedings{yan-murawaki-2025-addressing,
  author    = {Ruiyi Yan and Yugo Murawaki},
  title     = {Addressing Tokenization Inconsistency in Steganography and Watermarking Based on Large Language Models},
  booktitle = {Proceedings of the 2025 Conference on Empirical Methods in Natural Language Processing},
  pages     = {7076--7098},
  publisher = {Association for Computational Linguistics},
  address   = {Suzhou, China},
  year      = {2025},
  doi       = {10.18653/v1/2025.emnlp-main.361}
}

@article{yang2019rnnstega,
  author  = {Zhongliang Yang and Xiaoqing Guo and Zi-Ming Chen and Yongfeng Huang and Yu-Jin Zhang},
  title   = {{RNN-Stega}: Linguistic Steganography Based on Recurrent Neural Networks},
  journal = {IEEE Transactions on Information Forensics and Security},
  volume  = {14},
  number  = {5},
  pages   = {1280--1295},
  year    = {2019},
  doi     = {10.1109/TIFS.2018.2871746}
}

@article{yang2019tsrnn,
  author  = {Zhongliang Yang and Ke Wang and Jian Li and Yongfeng Huang and Yu-Jin Zhang},
  title   = {{TS-RNN}: Text Steganalysis Based on Recurrent Neural Networks},
  journal = {IEEE Signal Processing Letters},
  volume  = {26},
  number  = {12},
  pages   = {1743--1747},
  year    = {2019},
  doi     = {10.1109/LSP.2019.2920452}
}

@misc{yang2024qwen25,
  author        = {An Yang and Baosong Yang and Beichen Zhang and Binyuan Hui and Bo Zheng and Bowen Yu and Chengyuan Li and Dayiheng Liu and Fei Huang and Haoran Wei and Huan Lin and Jian Yang and Jianhong Tu and Jianwei Zhang and Jianxin Yang and Jiaxi Yang and Jingren Zhou and Junyang Lin and Kai Dang and Keming Lu and Keqin Bao and Kexin Yang and Le Yu and Mei Li and Mingfeng Xue and Pei Zhang and Qin Zhu and Rui Men and Runji Lin and Tianhao Li and Tianyi Tang and Tingyu Xia and Xingzhang Ren and Xuancheng Ren and Yang Fan and Yang Su and Yichang Zhang and Yu Wan and Yuqiong Liu and Zeyu Cui and Zhenru Zhang and Zihan Qiu},
  title         = {{Qwen2.5} Technical Report},
  year          = {2024},
  eprint        = {2412.15115},
  archiveprefix = {arXiv},
  primaryclass  = {cs.CL},
  url           = {https://arxiv.org/abs/2412.15115}
}

@inproceedings{zhang2021adg,
  author    = {Siyu Zhang and Zhongliang Yang and Jinshuai Yang and Yongfeng Huang},
  title     = {Provably Secure Generative Linguistic Steganography},
  booktitle = {Findings of the Association for Computational Linguistics: {ACL-IJCNLP} 2021},
  pages     = {3046--3055},
  publisher = {Association for Computational Linguistics},
  address   = {Online},
  year      = {2021},
  doi       = {10.18653/v1/2021.findings-acl.268}
}

@inproceedings{ziegler2019neural,
  author    = {Zachary Ziegler and Yuntian Deng and Alexander M. Rush},
  title     = {Neural Linguistic Steganography},
  booktitle = {Proceedings of the 2019 Conference on Empirical Methods in Natural Language Processing and the 9th International Joint Conference on Natural Language Processing},
  pages     = {1210--1215},
  publisher = {Association for Computational Linguistics},
  address   = {Hong Kong, China},
  year      = {2019},
  doi       = {10.18653/v1/D19-1115}
}

@inproceedings{chang2010paraphrases,
  author    = {Ching-Yun Chang and Stephen Clark},
  title     = {Linguistic Steganography Using Automatically Generated Paraphrases},
  booktitle = {Human Language Technologies: The 2010 Annual Conference of the
               North American Chapter of the Association for Computational Linguistics},
  pages     = {591--599},
  publisher = {Association for Computational Linguistics},
  address   = {Los Angeles, California},
  year      = {2010},
  url       = {https://aclanthology.org/N10-1084/}
}

@inproceedings{chang2012wordorder,
  author    = {Ching-Yun Chang and Stephen Clark},
  title     = {The Secret's in the Word Order: Text-to-Text Generation
               for Linguistic Steganography},
  booktitle = {Proceedings of {COLING} 2012},
  pages     = {511--528},
  publisher = {The COLING 2012 Organizing Committee},
  address   = {Mumbai, India},
  year      = {2012},
  url       = {https://aclanthology.org/C12-1032/}
}

@article{chen2011relative,
  author  = {Zhili Chen and Liusheng Huang and Wei Yang},
  title   = {Detection of Substitution-Based Linguistic Steganography
             by Relative Frequency Analysis},
  journal = {Digital Investigation},
  volume  = {8},
  number  = {1},
  pages   = {68--77},
  year    = {2011},
  doi     = {10.1016/j.diin.2011.03.001}
}

@article{xiang2018wordembedding,
  author  = {Lingyun Xiang and Jingmin Yu and Chunfang Yang
             and Daojian Zeng and Xiaobo Shen},
  title   = {A Word-Embedding-Based Steganalysis Method for Linguistic
             Steganography via Synonym Substitution},
  journal = {IEEE Access},
  volume  = {6},
  pages   = {64131--64141},
  year    = {2018},
  doi     = {10.1109/ACCESS.2018.2878273}
}
